\documentclass[prx,reprint,superscriptaddress]{revtex4-1}
\usepackage{graphics,epsfig,amsfonts,amssymb,amsmath,color}
\usepackage[normalem]{ulem}
\begin{document}

\newcommand{\hilight}[1]{\colorbox{yellow}{#1}}

\title{Donor Wavefunctions in Si Gauged by STM Images}
\author{A. L. Saraiva}
\affiliation{Instituto de F\'isica, Universidade Federal do Rio de Janeiro, Caixa Postal 68528, Rio de Janeiro, RJ 21941-972, Brazil}
\author {J. Salfi}
\affiliation{Centre for Quantum Computation and Communication Technology, School of Physics, The University of New SouthWales, Sydney, New SouthWales 2052, Australia}
\author {J. Bocquel}
\affiliation{Centre for Quantum Computation and Communication Technology, School of Physics, The University of New SouthWales, Sydney, New SouthWales 2052, Australia}
\author {B. Voisin}
\affiliation{Centre for Quantum Computation and Communication Technology, School of Physics, The University of New SouthWales, Sydney, New SouthWales 2052, Australia}
\author {S. Rogge}
\affiliation{Centre for Quantum Computation and Communication Technology, School of Physics, The University of New SouthWales, Sydney, New SouthWales 2052, Australia}
\author{Rodrigo B. Capaz}
\affiliation{Instituto de F\'isica, Universidade Federal do Rio de Janeiro, Caixa Postal 68528, Rio de Janeiro, RJ 21941-972, Brazil}
\author{M.J. Calder\'on}
\affiliation{Instituto de Ciencia de Materiales de Madrid, ICMM-CSIC, Cantoblanco, E-28049 Madrid (Spain)}
\author  {Belita Koiller}
\affiliation{Instituto de F\'isica, Universidade Federal do Rio de Janeiro, Caixa Postal 68528, Rio de Janeiro, RJ 21941-972, Brazil}

\date{\today}

\begin{abstract}
The triumph of effective mass theory in describing the energy spectrum of dopants does not guarantee that the model wavefunctions will withstand an experimental test. Such wavefunctions have recently been probed by scanning tunneling spectroscopy, revealing localized patterns of resonantly enhanced tunneling currents. We show that the shape of the conducting splotches resemble a cut through Kohn-Luttinger (KL) hydrogenic envelopes, which modulate the interfering Bloch states of conduction electrons. All the non-monotonic features of the current profile are consistent with the charge density fluctuations observed between successive $\{001\}$ atomic planes, including a counter-intuitive reduction of the symmetry -- a heritage of the lowered point group symmetry at these planes. A model-independent analysis of the diffraction figure constrains the value of the electron wavevector to $ k_0=(0.82\pm0.03)(2\pi/a_{\rm Si})$. Unlike  prior measurements, averaged over a sizeable density of electrons, this estimate is obtained directly from isolated electrons. We further investigate the model-specific anisotropy of the wave function envelope, related to the effective mass anisotropy. This anisotropy appears in the KL variational wave function envelope as the ratio between Bohr radii $b/a$. We demonstrate that the central cell corrected estimates for this ratio are encouragingly accurate, leading to the conclusion that the KL theory is a valid model not only for energies but for wavefunctions as well.
\end{abstract}
\maketitle

\section{Introduction}
\label{sec:introduction}

Modern applications of electronic quantum control at the single dopant level underline significant scientific challenges to the theory of doped semiconductors. Many challenges revolve around the determination of the donor electron wavefunction. In silicon, the rich conduction band structure renders the problem of hydrogenic impurities unsolvable even within the simple effective mass approximation. The pioneering work of Kohn and Luttinger (KL) provides a tentative answer within a variational framework~\cite{kohnPR1955a}. It is unclear, though, if the simplicity of effective mass approximation can withstand the comparison with the directly probed charge distribution of a donor. 

Furthermore, the variational principle guarantees that the minimal energy mean value is closest to the real ground state energy, but ascertains nothing about the optimal variational wavefunction. Variations with respect to the true ground state wavefunction only lead to second-order shifts in the ground state energy, meaning that even sizeable deviations from the ground state wavefunction can still lead to reasonably accurate energies. For example, energies estimated from the  KL trial function with two parameters are only less than 1\% higher than a model with four parameters by Kittel and Mitchell~\cite{kittel54}.

This indeterminacy represents a threat to the set of ideas that have led the pursuit of quantum technologies. To tame the electron is to manipulate its wavefunction. Entanglement between electrons is thought to be achievable by simply overlapping their wavefunctions, as to explore the Pauli exclusion principle and the spin-spin effective interaction that comes with it. Knowledge largely based on the effective mass KL model for dopant wavefunctions is the common root supporting all designs of donor-based quantum devices in Si~\cite{zwanenburg2013}. Should this wavefunction be sizeably wrong, the feasibility of most quantum devices would need to be revisited.

Recent scanning tunneling microscopy (STM) images of single donors near a silicon (001) surface revealed the theoretically predicted intricate interference patterns of such electronic states.~\cite{salfiNatMat2014} At first sight, these images do not communicate the simplicity of the KL wavefunction -- two kinds of non-trivial images are revealed, which we call here butterfly (B) or caterpillar (C). In Ref.~[\onlinecite{salfiNatMat2014}], these are referred to as types A and B, respectively. The B-type has a nodal line akin to a $p$ orbital symmetry, while the C-type has a symmetry closer to $s$. Both patterns are diagonally aligned and present mirror symmetry with respect to either  $[110]$ or $[1\overline{1}0]$ surface directions. The symmetry lines are consistently orthogonal to the dimer rows direction on the surface (which run along either direction in different terraces).

Here we reconcile the intrincate current profiles of the STM image with the simple KL theory by carefully considering the crystalline and electronic structures of bulk silicon. Perturbations by the STM tip, surface relaxation, reconstruction or passivation are not considered. The model is briefly reviewed in Sec.~\ref{sec:model}, and the notation for the atomic planes is discussed.  Section~\ref{sec:exp-model} juxtaposes theoretical and STM images for donors buried at different depths. In Sec.~\ref{sec:analysis}, we theoretically dissect the roles of the anisotropic mass, Bloch functions, and periodicity of the lattice, and thus pinpoint the main ingredients driving the anisotropy of the observed images.  The origin of the low-frequency interference patterns is detected in Sec.~\ref{sec:Fourier} analysing the charge distribution in Fourier space. This $k$-space analysis is further explored in order to estimate  $k_0$ and $b/a$ from the donor bound states in Sec.~\ref{sec:anisotropy}. Section~\ref{sec:conclusion} is devoted to our conclusions and final remarks.

\section{Model}
\label{sec:model}

\begin{figure*}
\includegraphics[clip,width=0.9\textwidth]{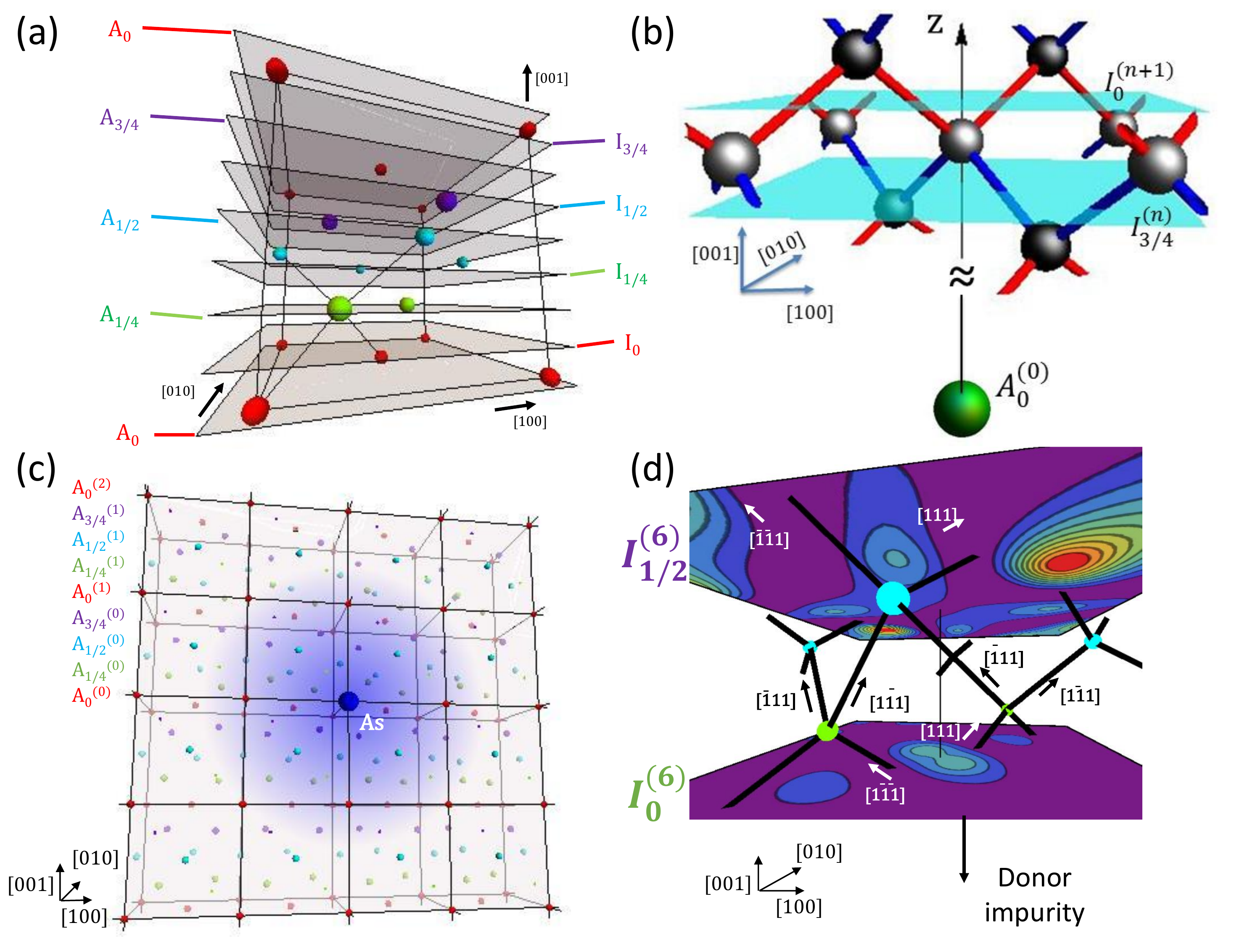}
\caption{(color online) Bulk crystal structure and notation adopted for atomic and interstitial $(001)$ planes. (a) Diamond structure sites represented by dots; an FCC cubic unit cell is also shown. The four inequivalent atomic planes $A_0$, $A_{1/4}$, $A_{1/2}$, and $A_{3/4}$, are indicated. Each interstitial plane is labelled as the atomic plane immediately below it, that is, $I_0$, $I_{1/4}$, $I_{1/2}$, and $I_{3/4}$. (b) Expanded view of the $A_0^{(n+1)}$ plane showing nearest-neighbors bonds and interstitial planes across them. The red bonds (M-shaped) and the blue ones (W-shaped) form zig-zag paths. The donor site and the symmetry axis $z$ (see text) are indicated. (c) Overall view of the crystal structure with a substitutional donor impurity located at the $A_{0}^{(0)}$ plane. Each cube here is a replica of the one in frame (a). (d) Details of the charge distributions of a combination of Bloch functions pinned at the donor site. Cuts are shown at the interstitial planes $I_{0}^{(6)}$ and $I_{1/2}^{(6)}$. One may identify shapes anticipating the B and C forms observed experimentally.}
\label{fig:crystal-structure}
\end{figure*}

Our model is formulated in the context of well established bulk Si structural and electronic properties. Silicon crystallizes in the diamond structure which consists of two interpenetrating face centered cubic (FCC) lattices shifted by $1/4$ of the cubic conventional cell body diagonal. All lengths referring to atomic positions and distances are given here in units of  the cubic cell lattice parameter $a_{\rm Si}= 0.543$ nm.

Fig.~\ref{fig:crystal-structure}(a)~shows a unitary FCC cube where   atomic positions of the diamond structure are indicated. We identify the stacking of $4$ inequivalent (001) atomic planes at heights $z = 0,1/4, 1/2, 3/4$ which we call $A_0, A_{1/4}, A_{1/2}$ and $A_{3/4}$ or, in general $A_{j/4}$ with $j=0,1,2,3$. In the perfect crystal, the next plane above, at $z= 1$ (also in the figure) is equivalent to the $A_0$ plane. It is convenient to label planes  intercalated midway between consecutive $A-$planes. We call  $I_{j/4}$ the intercalated plane  a distance $1/8$ above  $A_{j/4}$.

In the event that a substitutional donor (which defines the coordinates origin) is located a distance $d$ from a (001) surface at $z=d$,  translational symmetry is lost. We then refer  to a general atomic plane at  $z=n+j/4$  as $A_{j/4}^{(n)}$ with the donor at the $A_0^{(0)}$ plane.  Consistently,  intercalated planes at $z=n+j/4+1/8$ are called  $I_{j/4} ^{(n)}$.

Nearest neighbor (consecutive) atomic planes belong to different FCC sublattices.
The tetrahedral bonds for atom pairs  (one in each plane) define zig-zag paths crossing a single $I-$plane, as illustrated in Fig.~\ref{fig:crystal-structure}(b). Note that bonds across  $I_{j/4} ^{(n)}$ for j=0 or 2 are overall aligned with the $[110]$ diagonal while for $j = 1$ or 3 the paths  follow the $[1\overline{1}0]$ diagonal. This applies to all $j$ and  $(n)$ planes, including the surface $A_{j/4}^{(d-j/4)}$ at $z=d$, thus giving rise to two possible directions for the atomic reconstruction (dimerisation) on a Si (001) surface. In our theoretical analysis, we infer the surface dimer rows  orientation, which is the same as the zig-zag overall direction, according to the value of $j = 4 (d-n)$ with $n$ integer and $j=0, 1, 2, 3$.

Fig.~\ref{fig:crystal-structure}(b)  also illustrates that the symmetry of the $A_{j/4}^{(n)}$ plane depends on the sublattice to which this plane belongs.
For $j=0,2$ -- indices related to the same sublattice as the donor -- the planes show fourfold symmetry while those at the other sublattice  $j=1,3$  have a lower, twofold symmetry. Note that the $z$-axis crosses an atomic site only if it belongs to a $j=0$ atomic plane.

Electronic properties of Si are characterized by a band structure with sixfold degenerate conduction band minima (valleys), located at ${\bf k}_\mu$ along the $\langle 100 \rangle$  directions, $\mu =\pm x,\pm y, \pm z$. The valleys are anisotropic, \textit{i.e.}, they present different longitudinal and transverse effective masses ($m_\parallel =0.9163 m_e$ and $m_\perp=0.1905 m_e$ respectively~\cite{hensel1965}). Early measurements~\cite{feher1959} estimated the wavevector of the minima as $|{\bf k}_\mu|=k_0=(0.85 \pm 0.03) \left({\frac{2\pi}{a_{\rm Si}}}\right)$. Our measurements allow a less model-dependent, more accurate estimate of $k_0$, see Sec.~\ref{sec:anisotropy}.

The presence of a substitutional donor breaks the translational symmetry. In terms of electronic structure, a simple and successful description of shallow donors in Si was presented by Kohn and Luttinger within effective mass theory (EMT)~\cite{kohnPR1955a}.
The singular donor potential couples different valleys, leading to a non-degenerate ground state which involves a symmetric combination of the six valleys. The donor ground state variational wavefunction proposed by KL has the correct A$_1$-symmetry and is written in terms of envelopes and Bloch functions for each conduction band minimum

\begin{equation}
\Psi ({\bf r}) ={\frac{1}{\sqrt{6}}} \sum_\mu F_\mu ({\bf r}) e^{i {\bf k}_\mu \cdot {\bf r}} u_\mu (\bf r),
\label{eq:wave-function}
\end{equation}
where $F_\mu (\bf r)$ are envelope functions and $u_\mu (\bf r)$ are the periodic parts of the Bloch functions, given explicitly in Ref.~[\onlinecite{saraiva2011}] as obtained within first principles density functional theory. As a consequence of the mass anisotropy, the envelope functions have the shape of a deformed 1s orbital
\begin{equation}
F_{\pm z} ({\bf r}) ={\frac{1} {\sqrt{\pi a^2 b}}} \, \exp\left({-\sqrt{{\frac{x^2+y^2} {a^2}}+{\frac{z^2} {b^2}}}}\right)
\label{eq:envelopes}
\end{equation}
and similarly for the $x$ and $y$ valleys.

The effective Bohr radii $a$ and $b$ were calculated variationally by KL for a Coulomb donor potential. Additionally we take into account a central cell correction potential which is donor species dependent and chosen to reproduce each experimental ground state energy.~\cite{saraivaJPCM2015} Notice that the phenomenological central cell has a different radius $r_{cc}$ for the anisotropic model of the mass compared to the spherical mass model adopted in Ref.~[\onlinecite{saraivaJPCM2015}]. For the anisotropic model, the radii are $r_{cc}(\rm P)=120$ pm and $r_{cc}(\rm As)=128$ pm for P donors and As donors, respectively.

For P donors we get variational radii $a=1.13$~nm and $b=0.60$~nm, while for As donors we get $a=0.86$~nm and $b=0.46$~nm. Consequently, the envelope anisotropy is $b/a=0.53$ irrespective of the donor species. An isotropic approximation for the envelope $\exp(-r/a)$ leads to a single average Bohr radius $a=1.106$~nm for P and $a=0.815$~nm for As.~\cite{saraivaJPCM2015}
The exact value of $a$ and $b$ have an impact only on the size of the image, not on the details of its symmetry and oscillations. We adopt from now on $a=0.90$ nm and $b=0.52$ nm for the theoretical calculations (which lead to the same $b/a$ ratio as the original KL theory but incorporates the radii reduced by the central cell).

Our simulated STM images are generated using the following conditions: (i) Tersoff-Hamann approximation,~\cite{tersoffPRB1985} in which the tunneling current is proportional to the local density of states integrated over the bias energy window; (ii) constant-height mode; and (iii) the assumption that the electronic density associated to the defect level is not substantially modified by the presence of the surface, meaning also that the effects of surface reconstruction and hydrogen saturation are disregarded. Assumption (iii) seems rather drastic, and its validity can only be assessed {\it a posteriori}, by direct comparison to the experimental STM images.  A justification to attempt it comes from the agreement between tight-binding results and experiments in Ref.~[\onlinecite{salfiNatMat2014}], where the former indicates that valley populations for donors $>2.5$~nm from the surface differ from the bulk values by less than $5$\%. We choose the STM tip height to correspond to a distance $\delta$ above the surface atomic layer
 \begin{equation}
 \rho_d (x,y) = | \Psi(x, y, d+\delta)|^2 ,
\label{eq:rho}
\end{equation}
with $\delta = 1/8$, i.e. halfway between atomic planes and coincident with the intercalated planes $I^{(n)}_{j/4}$. Physically, this is justified by the fact that, in our bulk-truncated model, the STM tip would promote tunneling to/from the evanescent tail of dangling bonds. Therefore, cutting through the Si-Si bonds (at halfway between atomic planes) should provide a better description of the spatial dependence of the tunneling current in comparison, for instance, to cutting through atomic planes. As we shall see, by comparison with experimental images in Sec.~\ref{sec:exp-model}, this is precisely the case.

\begin{figure}
\leavevmode
\includegraphics[clip,width=0.45\textwidth]{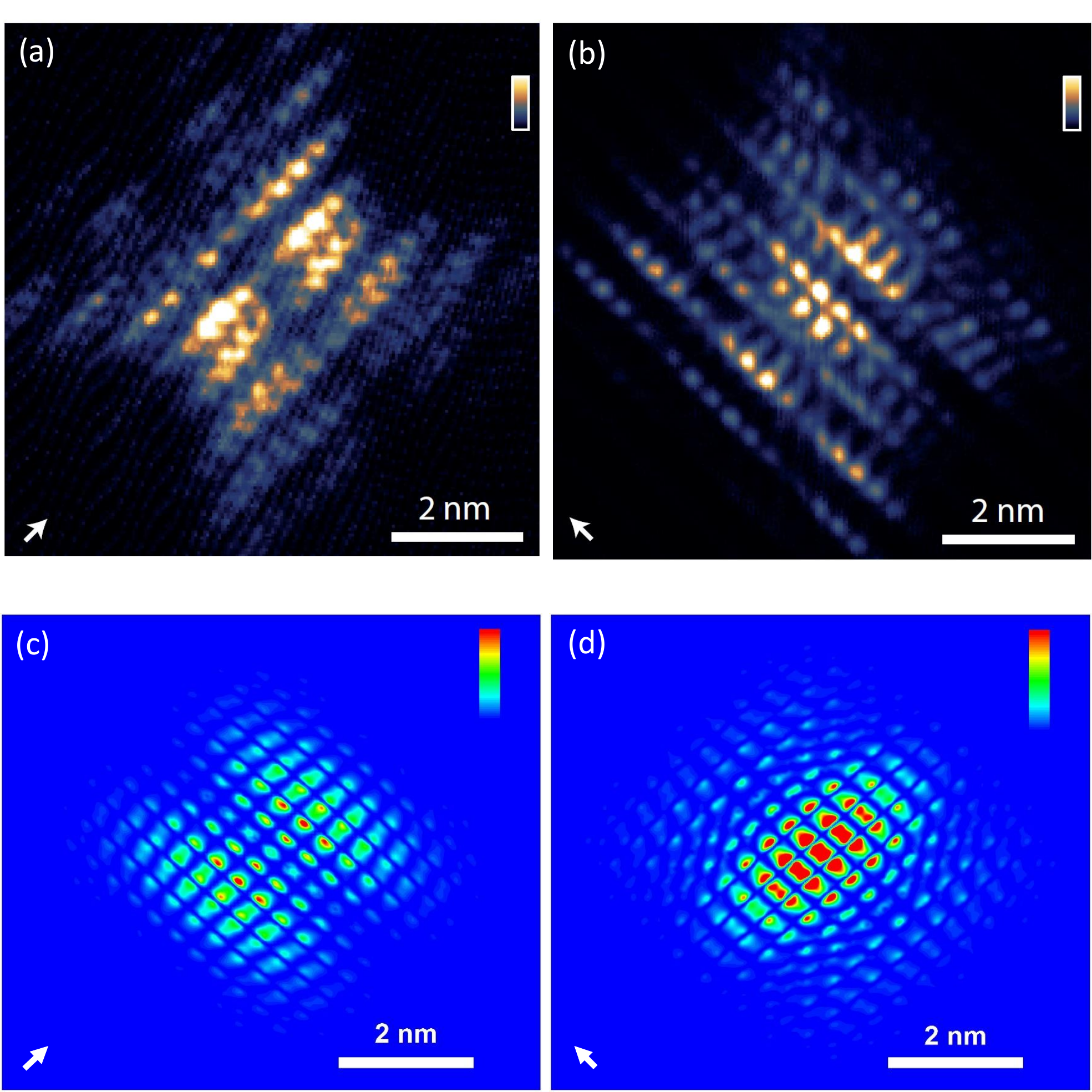}
\caption{Top: STM real space images of different donors showing the butterfly B (a) and caterpillar C (b) shapes. Bottom: images for the charge density from the wavefunction of donors at (c) $I_{3/4}^{(5)}$ at $z=5.875 \,a_{\rm Si}$ and (d) $I_{1/2}^{(5)}$ at $z=5.625 \,a_{\rm Si}$. The color scale in figures (c) and (d) ranges from 0 to $2\times 10^{-4} (a_{\rm Si}^{-3})$. The arrows indicate the surface dimer direction.}
\label{fig:EMT-vs-STM}
\end{figure}

\section{Real space results}
\label{sec:exp-model}

Measurements were performed as described in Ref.~[\onlinecite{salfiNatMat2014}]. Briefly, we directly measured the electronic ground state of shallow donors buried several nm underneath hydrogen-terminated (100) surfaces, by scanning tunneling spectroscopy (STS) in ultra high vacuum (UHV). Samples with either P or As donors were prepared and measured. The As-doped samples contained donors located at random depths\cite{salfiNatMat2014}.  Samples with P donors were fabricated by submonolayer PH$_3$ dosing and well-defined encapsulation by epitaxial silicon, using a procedure similar to the one in Ref.~[\onlinecite{miwaAPL2013}], but with an n-type substrate to promote elastic resonant transport~\cite{salfiNatMat2014,voisinJPCM2015l}. Experiments on the donors were carried out with atomic resolution in real space, in the single-electron transport regime with both the tip and a heavily-doped region of the sample, at liquid Helium temperatures, acting as transport reservoirs~\cite{molPRB2013}. Both the As and P donors were determined to be in a lightly doped $\sim 10$ nm thick layer of silicon\cite{salfiNatMat2014}, with nominally one donor per $30$ nm $\times$ $30$ nm surface area.  Analysis of $dI/dU$ lineshapes, where $U$ is the sample bias and $I$ is the tunneling current, revealed that the donors were typically coupled weakly to a buried reservoir, relative to the thermal energy of excitations in the source and drain~\cite{voisinJPCM2015l} ($k_BT=0.36$ meV).

To obtain high-resolution images of the neutral donor (D$^0$) orbitals we adopted an unconventional scheme employing an open-loop current measurement in the gap.  The first scan maps the topography of the hydrogen-terminated (100) surface at $U=-1.45$ V, revealing dimerization along [110] directions.  A second scan is made, following the topography of the first scan, but at a bias $U$ such that only the lowest energy D$^0$ state is in the bias window ($U_2=-0.8$ V).  This scheme allows us to routinely image with very high resolution the donor orbitals, whose surface electronic density can occupy more than $10$ nm $\times$ $10$ nm, with no cross-talk from the bulk states, or the two-electron (D$^-$) localized state which is found at $U\approx-1.1$ V sample bias.~\cite{salfiNatMat2014,voisinJPCM2015l}

The top panels of Fig.~\ref{fig:EMT-vs-STM} represent the two typical STM images for donors as obtained in our second scan measurements, with the dimer direction from the first scan indicated by an arrow. Both are twofold symmetric. Because of their general shapes they are named here butterfly (B) [Fig.~\ref{fig:EMT-vs-STM}(a)] and caterpillar (C) [Fig.~\ref{fig:EMT-vs-STM}(b)]. The butterfly is characterized by a nodal line in the low-frequency probability density that is always perpendicular to the dimer rows direction, while in contrast, the caterpillar is characterized by an antinodal line perpendicular to the dimer.

\begin{figure}
\leavevmode
\includegraphics[clip,width=0.45\textwidth]{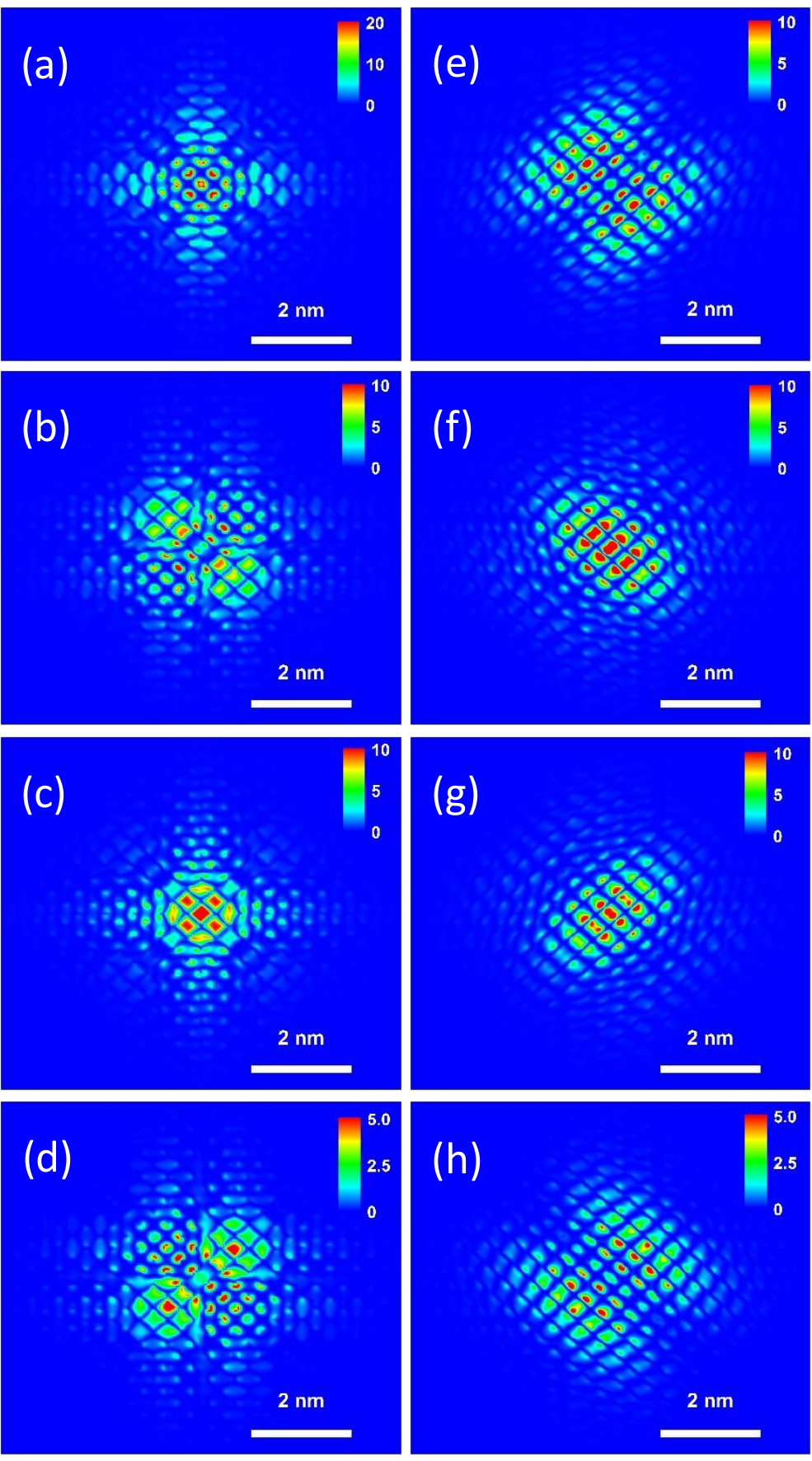}
\caption{Tomography of the KL wavefunction. Left: Cuts through atomic planes (a) $A^{(4)}_{0}$, (b) $A^{(4)}_{1/4}$, (c) $A^{(4)}_{1/2}$ and (d) $A^{(4)}_{3/4}$. Right: Interstitial cuts (e) $I^{(4)}_{0}$, (f) $I^{(4)}_{1/4}$, (g) $I^{(4)}_{1/2}$ and (h) $I^{(4)}_{3/4}$. The color code represents $|\Psi(z=z_0)|^2$ in units of $10^{-4} (a_{\rm Si}^{-3})$. The charge distributions calculated at the interstitial planes compare very well with the STM images, see Sec.~\ref{sec:exp-model} for details.}
\label{fig:all-planes}
\end{figure}

We now contrast the experiments with results from the model described in the previous section. Figs.~\ref{fig:all-planes}~(a) to (d)  show a sequence of inequivalent plane images calculated from Eq.~(\ref{eq:rho}) with $\delta = 0$, \textit{i.e.}, planes $A_{j/4}^{(4)}$ with $j=0,\ldots, 3$. These correspond to each of the four inequivalent atomic planes at $d\ge4$, while panels (e) to (h) show the four inequivalent interstitial ($\delta = 1/8$) cuts $I_{j/4}^{(4)}$ with $j=0,\ldots, 3$. Panel (a) is very similar to the well established behavior of the charge density cut for the (001) atomic plane containing the donor (see, for example, Fig. 2(d) in Ref.~[\onlinecite{koillerPRB2004}]).

Panels (a) and (c) in Fig.~\ref{fig:all-planes} correspond to atomic planes belonging to the same sublattice as the donor. The cut images display fourfold symmetry (cross-like) around the impurity projected position. The atomic planes in the sublattice not containing the donor lead to lower symmetry patterns (twofold, checkerboard-like). None of the atomic planes reproduce the STM images in the top panels of Fig.~\ref{fig:EMT-vs-STM}.

In contrast, calculated charge distributions at interstitial planes reveal very clear similarity to experimentally observed patterns (see Fig.~\ref{fig:EMT-vs-STM}). The experimentally measured images in the upper panels compare very well with the theoretically generated images for interstitial planes $I_{1/4}^{(5)}$ and $I_{3/4}^{(5)}$ in the lower panels. These are representative of B and  C patterns, respectively. In general, the butterfly shape is found for interstitial cuts $I_{j/4}^{(n)}$ for  $j=0$ or $3$, \textit{i.e.}, just above or below an $A_0^{(m)}$ plane. The caterpillar shape corresponds to images at $I_{j/4}^{(n)}$ for $j=1$ or $2$. The comparison between experimental data and theoretical calculations is compelling, including the orientation of the charge distribution with respect to the surface dimer rows (determined by the direction of the dangling bonds, but not explicitly accounted for in theory).

The calculated images corresponding to $n=5$  and $n=4$ $I$-planes (not shown) also manifest close similarities, but only on the overall shape (B or C). There are differences between the rapid oscillatory patterns, as expected since the complete image is not periodic along $z$.

We should stress that deviations from s-like charge densities have been observed for holes bound to Mn dopants in GaAs crystals in the form of a localized bowtie-shaped density.~\cite{yakuninPRL2004} This is a consequence of orbital hybridization orbital hybridization in the KL impurity model. Such an effect is generic to the valence band; the signature of d-orbital hybridization can also be seen for boron acceptors on the [100] surface of silicon\cite{molAPL2015}. We show here that the origin of the unusual low symmetry of the images in Ref.~[\onlinecite{salfiNatMat2014}] is the true interference of valleys, not an edge effect.

 \begin{figure}
\leavevmode
\includegraphics[clip,width=0.48\textwidth]{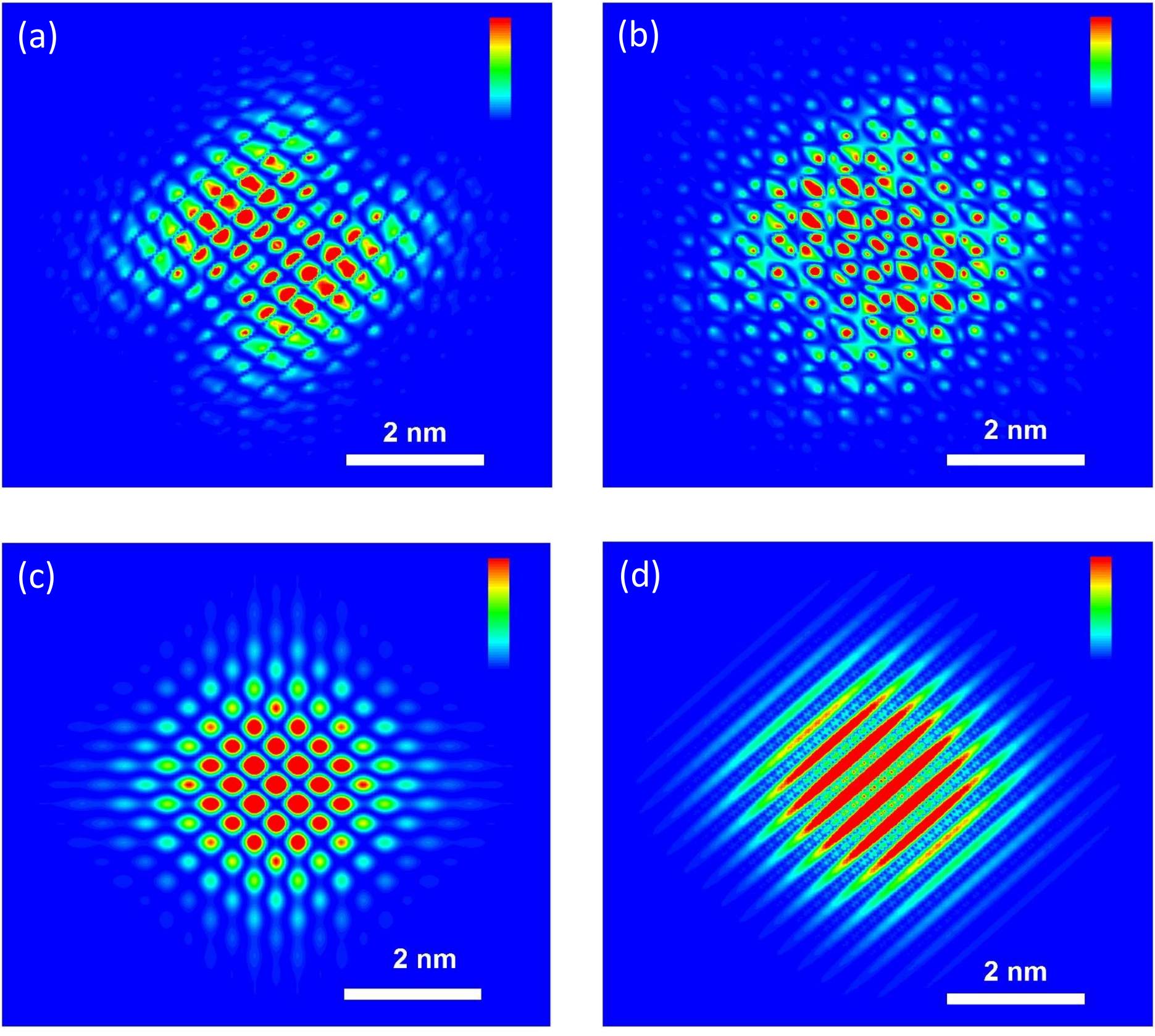}
\caption{Comparison between approximations for the charge distributions at plane $I^{(5)}_{0}$, \textit{i.e.}, at $z=5.125 a_{\rm Si}$. (a) Full wavefunction as described in Eq.~\ref{eq:wave-function} with anisotropic mass. (b) Full wavefunction with spherical (isotropic) mass. (c) Wavefunction with trivial periodic part $u_{\mu}=1$. (d) Wavefunction for $k_\mu=0$ for all $\mu$. The twofold symmetry is maintained as far as the periodic part $u_{\mu}$ is kept. Equivalent results are obtained for the caterpillar shape. The color scale ranges from 0 to $2\times 10^{-4} (a_{\rm Si}^{-3})$.}
\label{fig:dissect-butterfly}
\end{figure}

\section{Understanding the observed features: Wavefunction dissection}
\label{sec:analysis}

An intriguing result, related to the anisotropy of the donor bound state images, is the cyclic sequence of butterfly (B) and caterpillar (C) shapes as the donor's depth is varied [see Fig.~\ref{fig:all-planes} (e-h)]: $\ldots$B($\bar 1$)C(1)C($\bar 1$)B(1)B($\bar 1$),C(1)C($\bar 1)B(1)\ldots$, where S($i$) means S-shape with axis aligned with [1$i$0]. In words, B and C shapes alternate every two interstitial cuts and the symmetry axis changes from [110] to [1$\bar 1$0] for successive images of the same shape.

In order to investigate which element of the wavefunction leads to this behavior, we compare the complete KL model with less accurate approximations.
Fig.~\ref{fig:dissect-butterfly}  shows cuts across an interstitial B-type plane, $I^{(4)}_{3/4}$, comparing the following cases: (a) The complete KL expression, Eqs.~(\ref{eq:wave-function}) and (\ref{eq:envelopes}), (b) isotropic effective mass, simulated by taking $a = b=0.815$ nm in Eq.~(\ref{eq:envelopes}), (c) $u_\mu=1$, thus  disregarding the periodic parts of the Bloch functions, or (d) eliminating  the  plane-wave parts $e^{i k_\mu \cdot r}$ in Eq.~(\ref{eq:wave-function}), as if $k_\mu = 0$ for all $\mu$.

Anisotropic charge distributions are obtained in all cases except for the $u_\mu=1$ case, where a fourfold symmetric image results. We conclude that the reduced symmetry of interatomic planes on a diamond structure determines the observed fingerprint of anisotropy, since it is the periodic $u$ that encodes these features.  The same qualitative result is found in C-type images. In other words, the twofold symmetry of the images comes from the lattice.

It is possible to understand qualitatively the STM figures by simply analysing the geometric structure of bulk Si,  see Fig.~\ref{fig:crystal-structure}.
The B and C shapes result from the interference between the six valley states, imposed by the A$_1$ ground state (see Eq.~\ref{eq:wave-function}). Precursor shapes are clearly identified if we analyse an A$_1$-symmetric superposition
\begin{equation}
\Psi_{\rm CBE} ({\bf r}) ={\frac{1}{\sqrt{6}}} \sum_\mu  e^{i {\bf k}_\mu \cdot {\bf r}} u_\mu (\bf r).
\label{eq:Bloch-functions}
\end{equation}
This is illustrated in Fig.~\ref{fig:crystal-structure}(d), where we plot the electronic density cuts $|\Psi_{\rm CBE} (x,y,z_I)|^2$  at interstitial planes for heights $z_I = 6+ 1/8$ ($I^{(6)}_0$) and $z_I=6+1/2+1/8$ ($I^{(6)}_{1/2}$).

The 90-degrees rotation of the C and B symmetry axis in successive interstitial planes reflects alternation in the orientation of the zigzag pattern of the Si-Si bonds they cross [Fig.~\ref{fig:crystal-structure}(b)]. Interestingly, the orientation of the zigzag pattern of the Si-Si dangling bonds in the unreconstructed surface plane also determines the orientation of the dimer rows in the reconstructed surface. Therefore, the orientation of surface dimerization and anisotropy of the donor STM image are completely correlated (although not related by causality -- they share a common cause). In fact, from Figs.~\ref{fig:crystal-structure}(b) and (d) we note that the B or C symmetry axis orientation is always perpendicular to the zigzag pattern of the Si-Si bonds, which coincide with the dimer direction.

As a final remark, we address the fast oscillatory features in the charge distribution images. Both the plane-wave and the periodic parts contribute to charge oscillations as observed in Fig.~\ref{fig:dissect-butterfly}. The plane-wave components are the least intuitive, leading to oscillations that are incommensurate with the lattice, as discussed in Ref.~\cite{salfiNatMat2014}. The envelopes simply confine the interference pattern around the donor position and do not contribute to the oscillations.

Among the possible crystal orientations, [001] suffers from the least stringent oscillations of all directions, as pointed out in numerous works~\cite{koillerPRL2001, gamble2015}, resulting in the well determined cyclic sequence for (001) cuts in the donor charge distribution, following the lattice structure symmetry. This is related to the six phase factors $e^{i\mathbf{k}_{\mu} \mathbf{r}}$ for this particular orientation of the lattice. In a general direction, the periodicity is lost, see Appendix~\ref{sec:111}.

\section{Understanding the observed features in the Fourier space}
\label{sec:Fourier}
\begin{figure}
\includegraphics{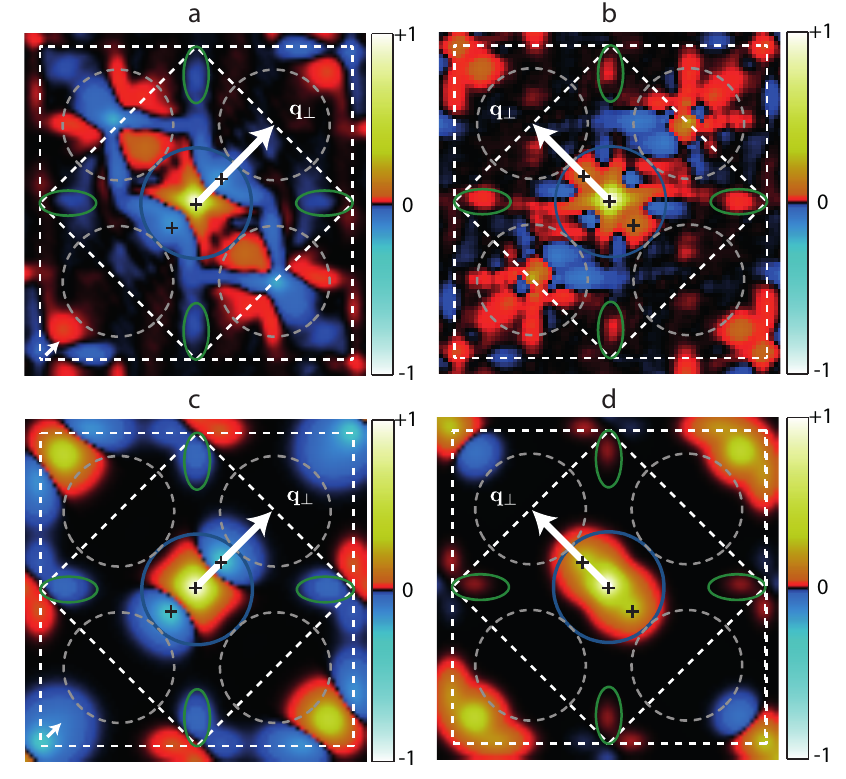}
\caption{Top: STM fourier space images of different donors showing the butterfly B (a) and caterpillar C (b) shapes. The four corners of the outermost dashed square boundary are reciprocal lattice vectors $2\pi(\pm1,\pm1)/a_{\rm Si}$. Bottom: images for the theoretical effective mass wavefunction of donors located at $z=0$ as seen from the interstitial planes (c) $I_0^{(5)}$ at $z=5.125 a_{\rm Si}$ and (d) $I_{1/4}^{(8)}$ at $z=8.375 a_{\rm Si}$.}
\label{fig:EMT-vs-STM-Fourier}
\end{figure}

In this section, we use KL equations~(\ref{eq:wave-function})-(\ref{eq:envelopes}) to construct an analytic model describing quantum interference processes in the donor probability density $|\Psi(\mathbf{r})|^2$.  These interference processes ultimately originate from linear superpositions of lattice-incommensurate valley wavevectors $\mathbf{k}_\mu$ and lattice commensurate reciprocal lattice vectors $\mathbf{G}$.  The model provides additional insight into the sequence B($\bar 1$)C(1)C($\bar 1$)B(1) in the interstitial planes $I_{j/4} ^{(n)}$ for $j=0,1,2,3$, and explains the origin of the sequence's lattice periodicity.  Moreover, the model forms the basis for the analysis in Section \ref{sec:anisotropy} where we extract quantitative information about $k_0$ and $b/a$ from the donor bound states.

Writing $u_\mu(\mathbf{r})=\sum_{\mathbf{G}}A_{k_\mu,G}\exp(i \mathbf{G}\cdot\mathbf{r})$, we find that $|\Psi(\mathbf{r})|^2$ can be written as,
\begin{align}
\nonumber|\Psi(\mathbf{r})|^2=\sum_{\mathbf{G}',\mathbf{G},\mathbf{k}_\mu,\mathbf{k}_\mu'} \Big[&A^*_{\mathbf{k}_\mu,\mathbf{G}}A_{\mathbf{k}_{\mu'},\mathbf{G}'}F^*_{\mu}(\mathbf{r})F_{\mu'}(\mathbf{r})\times\\&e^{i(-\mathbf{k}_{\mu}+\mathbf{k}_{\mu'}-\mathbf{G}+\mathbf{G}')\cdot\mathbf{r}}\Big],
\label{eq:fourier}
\end{align}
that is, slowly varying envelopes modulating oscillatory functions with spatial frequencies $\mathbf{q}=-\mathbf{k}_{\mu}+\mathbf{k}_{\mu'}-\mathbf{G}+\mathbf{G}'$.  Here the summation is over $-\mathbf{G}$ and $-\mathbf{k}_\mu$ in $\Psi^*(\mathbf{r})$ and $\mathbf{G}'$ and $\mathbf{k}_\mu'$ in $\Psi(\mathbf{r})$.  Enumerating $\mathbf{k}_\mu$, $\mathbf{k}_{\mu'}$ and $-\mathbf{G}+\mathbf{G}'$ in Table \ref{tab:valley-interference}, we identify the associated two-dimensional spatial frequencies $\vec{q}=(q_x,q_y)$.

\begin{table}
\caption{Mapping of plane-wave components in $\Psi(\mathbf{r})$ and $\Psi^*(\mathbf{r})$, to two-dimensional spatial frequencies $\vec{q}$ in $\rho_d(x,y)$. }
\begin{ruledtabular}
\begin{tabular}{| l | l | l | l | l |}
class & $\mathbf{k}_\mu$ & $\mathbf{k}_{\mu'}$ & $-\mathbf{G}+\mathbf{G}'$ & $\vec{q}=(q_x,q_y)$ \\
& & & $(2\pi/a_{\rm Si})$ & $(2\pi/a_{\rm Si})$ \\
\hline
\hline
1 & $\pm \mathbf{k}_{\mu x}$ & $\mathbf{k}_{\mu}$ & $(0,0)$ & $(0,0)$\\
1 & $\pm \mathbf{k}_{\mu y}$ & $\mathbf{k}_{\mu}$ & $(0,0)$ & $(0,0)$\\
1 & $\pm \mathbf{k}_{\mu z}$ & $\mathbf{k}_{\mu}$ & $(0,0)$ & $(0,0)$\\
\hline
2 & $\mp \mathbf{k}_{\mu x}$ & $+\mathbf{k}_{\mu z}$ & $(0,0)$ & $\vec{q}_{ex\pm}=(\pm0.85,0)$ \\
2 & $\mp \mathbf{k}_{\mu y}$ & $+\mathbf{k}_{\mu z}$ & $(0,0)$ & $\vec{q}_{ey\pm}=(0,\pm0.85)$\\
2 & $+\mathbf{k}_{\mu z}$    & $-\mathbf{k}_{\mu z}$    & $(0,0)$ & $(0,0)$\\
\hline
3 & $\pm \mathbf{k}_{\mu x}$ & $\mp \mathbf{k}_{\mu x}$ & $(\pm 2,0)$ & $\vec{q}_{px\pm}=(\pm0.3,0)$ \\
3 & $\pm \mathbf{k}_{\mu y}$ & $\mp \mathbf{k}_{\mu y}$ & $(0,\pm 2)$ & $\vec{q}_{py\pm}=(0,\pm0.3)$ \\
3 & $\pm \mathbf{k}_{\mu x}$ & $\pm\mathbf{k}_{\mu y}$ & ($\pm1,\mp1)$ & $\vec{q}_{k\pm}=\pm (0.15,-0.15)$\\
3 & $\pm \mathbf{k}_{\mu x}$ & $\mp\mathbf{k}_{\mu y}$ & ($\pm1,\pm1)$ & $\vec{q}_{b\pm}=\pm (0.15,0.15)$ \\
\end{tabular}
\end{ruledtabular}
\label{tab:valley-interference}
\end{table}

\begin{figure*}
\includegraphics{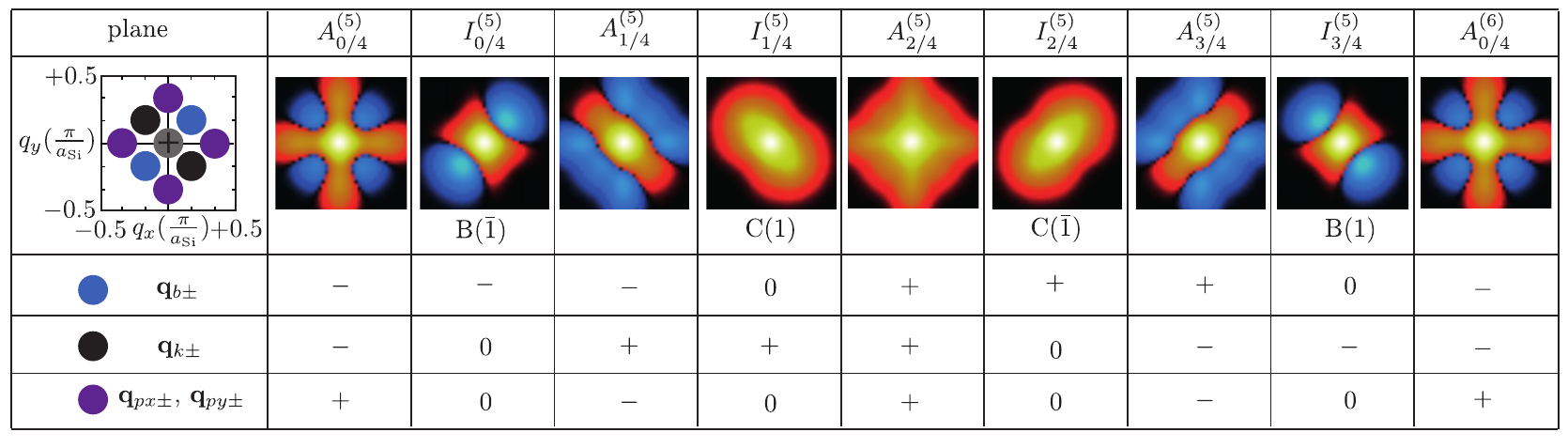}
\caption{Interference processes for the atomic and interstitial planes in the range $|q_x|\leq 2\pi/a_{\rm Si}$ and $|q_y|\leq 2\pi/a_{\rm Si}$. }
\label{fig:valley-interference-composite}
\end{figure*}

The main insight of Table~\ref{tab:valley-interference} is to classify the oscillatory frequencies in Equation~(\ref{eq:fourier}), and to identify the inter-valley scattering processes producing special spatial frequencies $\vec{q}$.  The first class in Table \ref{tab:valley-interference} is the set of trivial non-interfering terms with $\vec{q}=0$ which occur for $\mathbf{k}_\mu' +\mathbf{G}'=\mathbf{k}_\mu+\mathbf{G}$. The second class is the set of interference terms whose oscillatory frequency $\mathbf{q}$ does not have lattice periodicity $a_{\rm Si}$ in the $\mathbf{z}$ direction, due to the presence of $\pm\mathbf{z}$ valleys.  Spatial frequencies $\vec{q}_{ex\pm}=\pm2\pi(0.85,0)/a_{\rm Si}$ and $\vec{q}_{ey\pm}=\pm2\pi(0,0.85)/a_{\rm Si}$ belong to this category, and originate from cross-terms of $\mp\mathbf{x}$ and $\mp\mathbf{y}$ valleys, respectively, with  $\mathbf{z}$ and $-\mathbf{z}$ valleys.  Since $\mathbf{G}=\mathbf{G}'$, the terms listed in class 2 in Table~\ref{tab:valley-interference} correspond to interference due to scattering by the valley-orbit potential of the donor.

A third class in Table \ref{tab:valley-interference} is the set of interference terms involving the $\mathbf{x}$ and $\mathbf{y}$ valleys and $\mathbf{G}'\neq\mathbf{G}$.  These terms correspond to interference due to scattering between valleys produced by the sharp central-cell potential of the donor, and scattering by the lattice periodic potential.  Since these terms do not involve the $\mathbf{z}$ valleys, their oscillatory frequencies $\mathbf{q}$ have the lattice period $a_{\rm Si}$ in the $\mathbf{z}$ direction.  Spatial frequencies $\vec{q}_{px\pm}=2\pi(\pm 0.3,0)/a_{\rm Si}$ and $\vec{q}_{py\pm}=2\pi(0,\pm 0.3)/a_{\rm Si}$ belong to this category, which are cross-terms of $\pm \mathbf{x}$ and $\mp \mathbf{x}$ valleys, and cross-terms of $\pm \mathbf{y}$ and $\mp \mathbf{y}$ valleys, respectively.  Similarly, $\vec{q}_{k\pm}=\pm 2\pi(0.15,-0.15)/a_{\rm Si}$ and $\vec{q}_{b\pm}=\pm 2\pi(0.15,0.15)/a_{\rm Si}$ belong to this category, and are cross terms of $\mathbf{x}$ and $\mathbf{y}$ valleys.

We now turn to the real parts of the Fourier transforms of the data in Fig.~\ref{fig:EMT-vs-STM}(a) and Fig.~\ref{fig:EMT-vs-STM}(b), shown in Fig.~\ref{fig:EMT-vs-STM-Fourier}(a) and Fig.~\ref{fig:EMT-vs-STM-Fourier}(b) respectively.  As indicated in the colour scale, red-yellow (blue-cyan) denotes positive (negative) values. In both Fig.~\ref{fig:EMT-vs-STM-Fourier}(a) and Fig.~\ref{fig:EMT-vs-STM-Fourier}(b) we observe ellipse-like features in the vicinity of $\vec{q}_{ex\pm}$ and $\vec{q}_{ey\pm}$ and more complex features in the vicinity of $\vec{q}=0$, as reported in Ref.~[\onlinecite{salfiNatMat2014}].  The presentation of Fourier transforms herein differs compared to Ref.~[\onlinecite{salfiNatMat2014}] only by showing the real parts, rather than the absolute value.

Examining a large set of measured donors we find that the B and C symmetry charge densities differ in Fourier space along the direction $\vec{q}_\perp$ (Fig. \ref{fig:EMT-vs-STM-Fourier}) that we define as perpendicular to the dimer's rows primary spatial frequency.  For example, the B($\bar 1$) pattern in Fig.~\ref{fig:EMT-vs-STM-Fourier}(a) is negative at the $\vec{q}_{b+}=2\pi(0.15,0.15)/a_{\rm Si}$ side peak, positive at $\vec{q}=0$, and negative at the $\vec{q}_{b-}=2\pi(-0.15,-0.15)/a_{\rm Si}$ side peak.  In contrast, the C(1) pattern [Fig.~\ref{fig:EMT-vs-STM-Fourier}(b)] is positive at the $\vec{q}_{k+}=2\pi(0.15,-0.15)/a_{\rm Si}$ side peak, positive at $\vec{q}=0$, and again positive at the $\vec{q}_{k-}=2\pi(-0.15,0.15)/a_{\rm Si}$ side peak. Features within gray dashed circles centred at $\pi(\pm 1,\pm 1)/a_{\rm Si}$ are due to the $2\times1$ reconstruction.~\cite{salfiNatMat2014}

For comparison with theory we show the real part of the Fourier transforms for planes $I_{0/4} ^{(5)}$ and $I_{1/4} ^{(8)}$ in Fig.~\ref{fig:EMT-vs-STM-Fourier}(c) and Fig.~\ref{fig:EMT-vs-STM-Fourier}(d) respectively, which are B($\bar 1$) and C(1) patterns.  First we note that the features observed at $\vec{q}_{ex\pm}$ and $\vec{q}_{ey\pm}$ in the measurements are reproduced, while reconstruction-related features are absent, because of assumption (iii) in Section~\ref{sec:model}.  Focusing on the region $|q_x|,|q_y|\leq \pi/a_{\rm Si}$, the $\vec{q}_{b\pm}$ side peaks are negative for the B($\bar 1$) pattern, and the $\vec{q}_{k\pm}$ side peaks are positive for the C(1) pattern.   Consequently, the interstitial plane reproduces the main Fourier-space features classifying the B and C patterns in the measurements.

In rows 1 and 2 of Fig.~\ref{fig:valley-interference-composite}(a) we present in order the sequence of Fourier transforms alternating between atomic and interstitial for $n=5$.  First, we note that the inequivalence between $[110]$ and $[1\bar10]$ directions in the interstitial planes is reflected by alternation of the sign and amplitude of peaks at $\vec{q}_{b\pm}$ and $\vec{q}_{k\pm}$.  As expected from $\mathbf{q}$ in Table \ref{tab:valley-interference}, the peaks centred at $\vec{q}_{b\pm}$ and $\vec{q}_{k\pm}$ are periodic with the lattice constant $a_{\rm Si}$ in the $\mathbf{z}$ direction.  Peaks at $\vec{q}_{px\pm}$ and $\vec{q}_{py\pm}$ are absent in the interstitial plane, but represented in the atomic plane, and as expected from $\mathbf{q}$ in Table \ref{tab:valley-interference}, they vary twice as fast with $z$ compared with $\vec{q}_{b\pm}$ and $\vec{q}_{k\pm}$.

\section{Conduction Band Minima and Envelope Anisotropy}
\label{sec:anisotropy}

\begin{figure}
\includegraphics{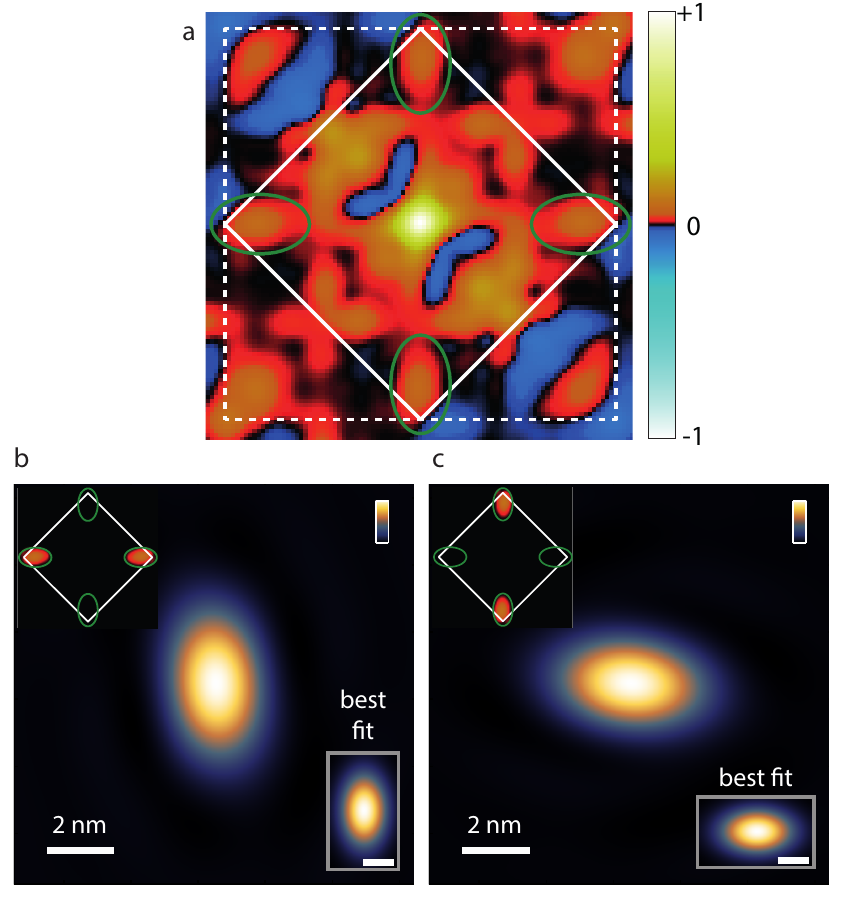}
\caption{Top: STM Fourier space image of a caterpillar (C) shape (a).  The four corners of the outermost dashed boundary are reciprocal lattice vectors $2\pi(\pm1,\pm1)/a_{\rm Si}$. Bottom: Fourier filtered data isolating the $x$ (b) and $y$ (c) ellipse features (top-left inset), real space data shifted back to the origin in Fourier space, and best fit to bulk-truncation model from the main text.}
\label{fig:ellipsefit}
\end{figure}

The anisotropy $b/a$ of the envelope functions and the wavevector $|\mathbf{k}_\mu|=k_0$ of the conduction band minima in KL equations~(\ref{eq:wave-function}) and (\ref{eq:envelopes}) can be determined from the spatial frequency and anisotropy of the ellipse-shaped features found in the Fourier space data at $\vec{q}_{ex\pm}=\pm\mathbf{x}k_0$ and $\vec{q}_{ey\pm}=\pm\mathbf{y}k_0$. Our analysis relies on the isolation of these features in Fourier space, which are highlighted for a P donor in Figure \ref{fig:ellipsefit}(a), and the specific origin of this interference term as identified in Table~\ref{tab:valley-interference}.

Owing to localization of the donor bound states, spatial oscillations at wavevectors $\mathbf{q}=-\mathbf{k}_{\mu}+\mathbf{k}_{\mu'}-\mathbf{G}+\mathbf{G}'$ in Table \ref{tab:valley-interference} and Equation \ref{eq:fourier} modulate slowly varying envelopes $A^*_{\mathbf{k}_\mu,\mathbf{G}}A_{\mathbf{k}_{\mu'},\mathbf{G}'}F^*_{\mu'}(\mathbf{r})F_\mu(\mathbf{r})$.  Focusing on the relevant $\mathbf{k}_\mu$, $\mathbf{k}_{\mu'}$, $\mathbf{G}$, and $\mathbf{G}'$ in Table \ref{tab:valley-interference}, we determine from Equation \ref{eq:fourier} that contributions to $|\Psi(\mathbf{r})|^2$ at frequencies $\vec{q}_{ex\pm}=\pm\mathbf{x}k_0$ and $\vec{q}_{ey\pm}=\pm\mathbf{y}k_0$ are of the form
\begin{align}
\label{eq:nex}
n_{ex}(\mathbf{r})&=C_x\cos(k_0z)\cos(k_0x)F_x(\mathbf{r})F_z(\mathbf{r})\textrm{, and}\\
\label{eq:ney}
n_{ey}(\mathbf{r})&=C_y\cos(k_0z)\cos(k_0y)F_y(\mathbf{r})F_z(\mathbf{r}).
\end{align}
For fixed $z=d$, $n_{e\mu}(\mathbf{r})$ is a product of envelope functions modulated by oscillations $\cos(\mathbf{k}_\mu\cdot\mathbf{r})$, with a constant pre-factor $C_\mu\cos(z_0d)$, where $C_\mu$ depends on the $A_{\mathbf{k},\mathbf{G}}$.

Our model for the conduction band minima, described in Appendix~\ref{sec:uncertainty}, relies only on the definite parity of the donor envelope functions, and allows for the extraction of $k_0$ separately for the $x$ and $y$ direction. Fitting the data in Figure \ref{fig:ellipsefit}(a) around $\vec{q}=\mathbf{x}k_0$ and $\vec{q}=\mathbf{y}k_0$ separately, we obtain essentially identical extrema $k_0 = (0.83 \pm 0.02)(2\pi/a_{\rm Si})$ for the $x$ and $y$ conduction band valleys, respectively.  The errors are dominated by inaccuracy in the determination of $\vec{q}$ relative to lattice frequencies $2\pi/a_{\rm Si}$, since the latter are blurred in topography due to finite sampling and instrumental errors, as discussed in Appendix~\ref{sec:uncertainty}.

The observed anisotropy of the Fourier representation of $n_{ex}(\mathbf{r})$ and $n_{ey}(\mathbf{r})$ in Fig.~\ref{fig:ellipsefit}(a) results from the anisotropy $b/a$ of $F_x(x,y,d)$ and $F_y(x,y,d)$ respectively, since $F_z(x,y,d)$ is isotropic in the $x-y$ plane. To determine $b/a$, we first isolated the ellipse-like features in Figure \ref{fig:ellipsefit}(a) corresponding to $n_{ex}(\mathbf{r})$ and $n_{ey}(\mathbf{r})$ in our model by applying a filter in Fourier space passing the relevant data inside the green boundaries.  Results of the filter for the $x$ and $y$ valleys are shown in the upper-left insets of Figure \ref{fig:ellipsefit}(b) and Figure \ref{fig:ellipsefit}(c), respectively.  Now isolated, these features can be directly fit to model expressions (\ref{eq:nex}) and (\ref{eq:ney}).

Independent of the details of the Fourier filter employed to isolate the features, we obtain envelope anisotropies $b/a=0.54\pm0.02$ and $b/a=0.53\pm0.02$ for the $x$ and $y$ valleys, respectively. The real space data are shown in Figure \ref{fig:ellipsefit}(b) and Figure \ref{fig:ellipsefit}(c) after shifting back to the origin in Fourier space using $k_0$ from above.  The anisotropy of this data reflects the anisotropy of the envelopes $F_x(\mathbf{r})$ and $F_y(\mathbf{r})$. The lower right insets of Figure \ref{fig:ellipsefit}(a) and \ref{fig:ellipsefit}(b) respectively shows the model calculation for the least-squares parameters, demonstrating excellent agreement with the data.

The results obtained for the donor in Figure \ref{fig:ellipsefit}(a) are representative of other donors.  Across a total of six donors (four As donors and two P donors), the values obtained for $k_0$ are very reproducible; best fit values for $k_0$ in the range  $[0.81, 0.84](2\pi/a_{\rm Si})$ are obtained, with similar confidence intervals between $0.02(2\pi/a_{\rm Si})$ and $0.03(2\pi/a_{\rm Si})$.  The mean value of all measurements is $k_0=(0.82\pm0.03)(2\pi/a_{\rm Si})$.  Across the same set of six donors, we obtain $b/a=0.49\pm0.03$ as the mean value for both the $x$ and $y$ valleys.  Nine of the twelve best fit values for $b/a$ fall in the range $[0.47, 0.58]$, though we also found three anomalously smaller values of $\sim 0.40 \pm 0.03$. The larger spread of $b/a$ values could be caused by disorder induced by the presence of randomly located dangling bonds on the surface.  We observe that these dangling bonds, whose position can be identified with atomic resolution, spatially modulate the density of states of the conduction and valence bands on length scales similar to the donor's effective Bohr radius.

\section{Discussions and Conclusion}
\label{sec:conclusion}

The notable resemblance between the conduction splotches and a cut of the KL wavefunction charge density entails two results. Firstly, that the KL theory is not limited to the estimation of the energetics of an electron bound to a dopant. It may be used as a good model of the electron wavefunction as well. Secondly, that the complex profile of the STM image is a surprisingly accurate depiction of the bulk wavefunction, and the surface does not disturb the electronic wavefunction beyond recognition.

From the first result, we may extrapolate important implications to the design of quantum devices. Theoretical proposals based on the KL model of the donor wavefunction should be reasonably accurate, as long as the fast oscillations of the wavefunction are consistently taken into account~\cite{gonzalez2014,saraivaJPCM2015,gamble2015}. Moreover, the resilience of the valley coherent interference and overall shape of the wavefunction reiterates that the donor-bound electron is robustly shaped by the donor potential and valley-orbit coupling, and disturbances like the passivated interface and the STM tip do not alter significantly the dopant wavefunction.

From the second result we conclude that the reduced symmetry of the STM images is not an artifact of the experiment, but reflects the true nature of the bulk donor wavefunction. Measurements made at the surface are good estimates of the bulk behavior. We may use that to appoint experimental values to the wavevector at the conduction band minima $k_0=|k_{\mu}|$ and the anisotropy $b/a$. 

While early reported measurements for $k_0$ vary from 0.77 $2\pi/a_{Si}$~\cite{macfarlanePR1955} to 0.85 $2\pi/a_{Si}$\cite{baldereschiPRB1970}, we obtain $k_0=(0.82\pm0.03)(2\pi/a_{\rm Si})$, which is independent of the model adopted for the envelope function of the dopant ground state. 

Adopting a KL envelope, we obtain a ratio $b/a=0.49\pm0.03$ that fits the experimental data. Most frequently, we measure ratios near 0.52, in excellent agreement with the theoretical values obtained within a central cell corrected model (0.53 for both As and P) and is slightly lower than the ratio obtained by KL without central cell (0.58 using modern values of the effective mass and dielectric constant). It is expected that $b/a$ changes as the wavefunction is probed at diferent distances from the impurity center. Near the center, the Schr\"odinger equation is dominated by the nearly spherically symmetric potential energy, pushing the ratio $b/a$ towards 1. Far from the center, however, the kinetic energy dominates at large, and one expects $b/a\approx\sqrt{m_\perp/m_\parallel}=0.46$. This could explain the fits from measurements being systematically lower than the theoretical estimates.

 Ideally, the anisotropy would be estimated both theoretically and experimentally as a function of the distance from the nucleus. Unfortunately, a trial variational wavefunction with too many fitting parameters leads to unreliable results, and the lack of a definite estimate of donor depth impairs this connection from the experimental point of view. A rigorous description should allow for the anisotropy to be a function of the distance from the nucleus. Unfortunately, within the variational scheme, this involves increasing the number of parameters in the trial envelope functions, with questionable results. At the current stage of Si quantum technologies~\cite{zwanenburg2013}, most applications do not urgently require such refinement.

The KL theory is shown here to be accurate and applicable to new quantum technologies -- a remarkable feat for a 60-year-old model.

This work is supported by the European Commission Future and Emerging Technologies Proactive Project MULTI (317707) and the ARC Centre of Excellence for Quantum Computation and Communication Technology (CE110001027), and in part by the US Army Research Office (W911NF-08-1-0527). ALS, RBC, and BK performed this work as part of the Brazilian National Institute for Science and Technology on Quantum Information and also acknowledge partial support from the Brazilian agencies FAPERJ, CNPq, CAPES. MJC acknowledges funding from MINECO (Spain) via Grant No. FIS2012-33521. ALS, BK and MJC acknowledge support from a bilateral CNPq (Brazil)- CSIC (Spain) grant.

\appendix

\renewcommand{\thefigure}{A\arabic{figure}}
\renewcommand{\thetable}{A\arabic{table}}
\setcounter{figure}{0}    

\section{Perspectives for sub-(111) surface donors} \label{sec:111}
\begin{figure}
\leavevmode
\includegraphics[clip,width=0.35\textwidth]{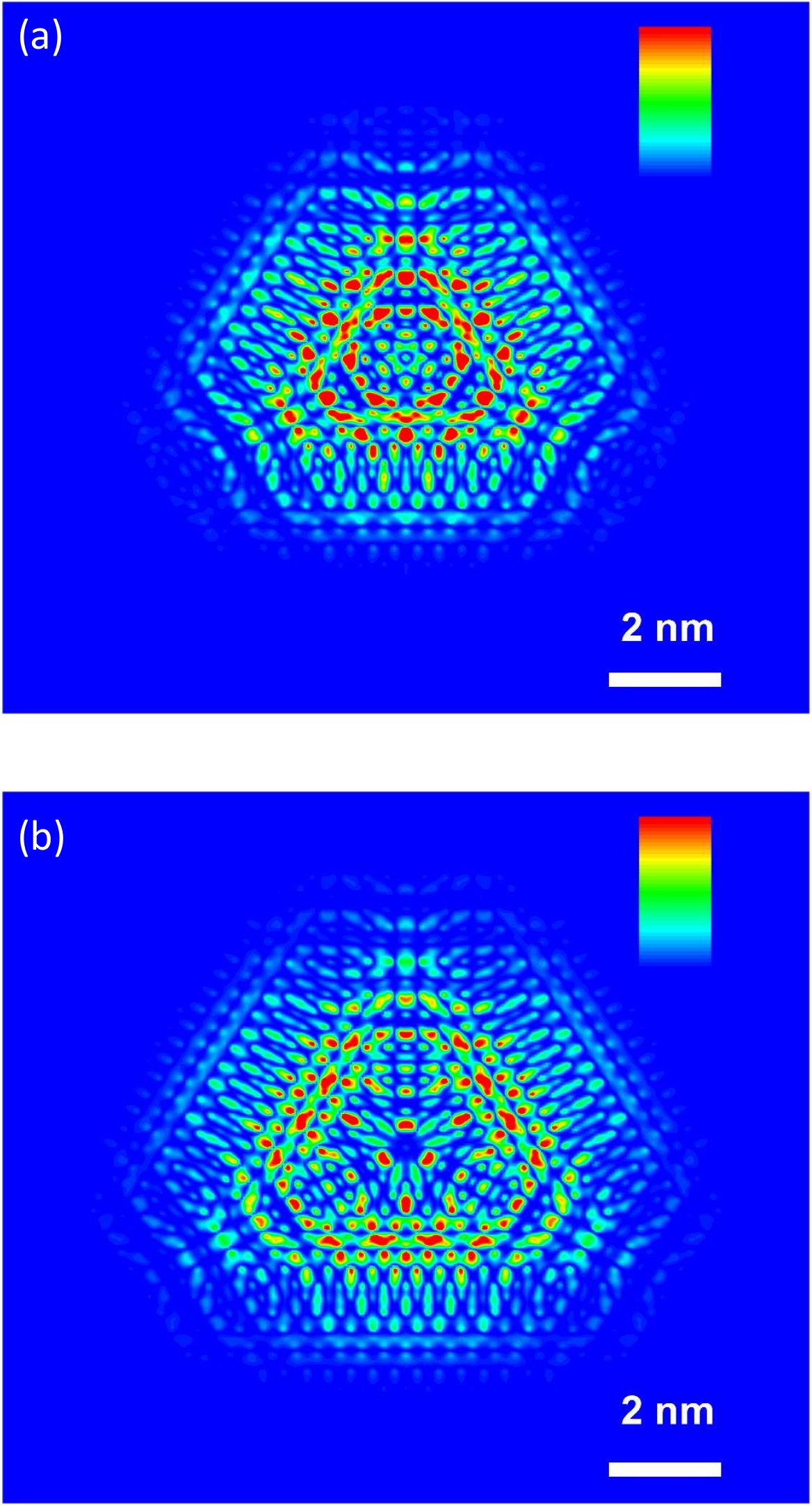}
\caption{Charge densities calculated at two structurally equivalent interstitial [111] plane at a distance $4.125\times\sqrt{3}a_{\rm Si}$ from the donor. The color scale ranges from 0 to $5\times10^{-6} (a_{\rm Si}^{-3})$ on both images. Note significant differences at the center and close similarity toward the edges due to the severe valley intereference along the $\langle 111\rangle$ direction.}
\label{fig:111}
\end{figure}

Experimental data presented herein focuses on the H/Si(100)2$\times$1 surface, prepared by UHV annealing to produce large defect free terraces whose dangling bonds are saturated by UHV exposure to atomic hydrogen.~\cite{bolandPRL1990}  While preparation of hydrogen terminated Si(110) surfaces is more difficult,~\cite{ouraSurfSciRep1999} the Si(111) surface presents another immediate possibility since large defect-free terraces of the well-known H/Si(111)7$\times$7 surface can be obtained by aqueous treatment in hydrofluoric acid.~\cite{dagataAPL1990} The $1\times 1$ reconstruction of H/Si(111) can also be obtained by wet chemical treatment,~\cite{higashiAPL1990} or in UHV by thermal treatment of the H/Si(111)7$\times$7 surface under flux of atomic hydrogen.~\cite{owmanSurfSci1994}

We present in this appendix images to be expected for STM of donors under surfaces  with (111) orientation. In analogy with the results for (001) surfaces, one might expect a cycle  of 3 inequivalent charge distributions at successive (111)  interstitial cuts, which is not obtained here. We observe all cuts with an overall triangular shape at large scales and a central part with no periodicity or similarity among different cuts for all donor depths we have examined. 

We attribute the absence of periodicity in the central region of the figures to the interference effects among the incommensurate plane waves at $k_\mu$ and the lattice-commensurate plane waves expansion of the periodic functions $u_\mu$. Interference effects are known to be stronger along (111) than (001) directions\cite{koillerPRL2001}. 

\section{Uncertainty in $k_0$} \label{sec:uncertainty}

Real space confinement  $\delta r\sim 3$ nm of surface charge density of the bound state blurs the Fourier space peaks by an amount $\delta q\sim 2\pi/\delta r\sim 0.2(2\pi/a_{\rm Si})$.  Nevertheless, $k_0$ can be determined in Fourier space with accuracy better than $\delta q$, as follows, because of the inversion symmetry of the donor envelope functions.  Expanding Equation [\ref{eq:nex}] around $\vec{q}=\mathbf{x}k_0$ in Fourier space, we obtain $n_{ex\pm}(q_x,q_y,z)=\sum_{n,m}M_{n,m}(q_x\pm k_0)^{n}q_y^{m}/n!m!$, where $M_{n,m}=\tfrac{1}{2}[\partial^{n+m} F_{xz}(q_x,q_y,z)/\partial q_x^{n}\partial q_y^{m}]|_{\vec{q}=0}$, and $F_{xz}(q_x,q_y,z)=\int\int dx dy e^{-i(q_xx+q_yy)}F_x(x,y,z)F_z(x,y,z)$.  Values for $M_{n,m}=\frac{1}{2}\int\int dxdy (-ix)^n(-iy)^m F_x(x,y,z)F_z(x,y,z)$ are non-zero when $n$ and $m$ are both even, when the integrand has even parity in $x$ and $y$ coordinates. To lowest non-zero order in $q$, we have $n_{ex}(q_x,q_y,z)=M_{0,0}-|M_{2,0}|(q_x \pm k_0)^2/2-|M_{0,2}|q_y^2/2$ in the vicinity of $\vec{q}=\mathbf{x}k_0$.  Following the same approach, we obtain $n_{ey}(q_x,q_y,z)=M_{0,0}-|M_{2,0}|(q_y \pm k_0)^2/2-|M_{0,2}|q_x^2/2$ in the vicinity of $\vec{q}=\mathbf{y}k_0$. In other words, $\mathbf{k}_\mu$ define the extrema in the Fourier space distributions of Equations [\ref{eq:nex}] and [\ref{eq:ney}].  

\section {Estimating the donor position}
\label{subsec:position}

\begin{figure}
\leavevmode
\includegraphics[clip,width=0.35\textwidth]{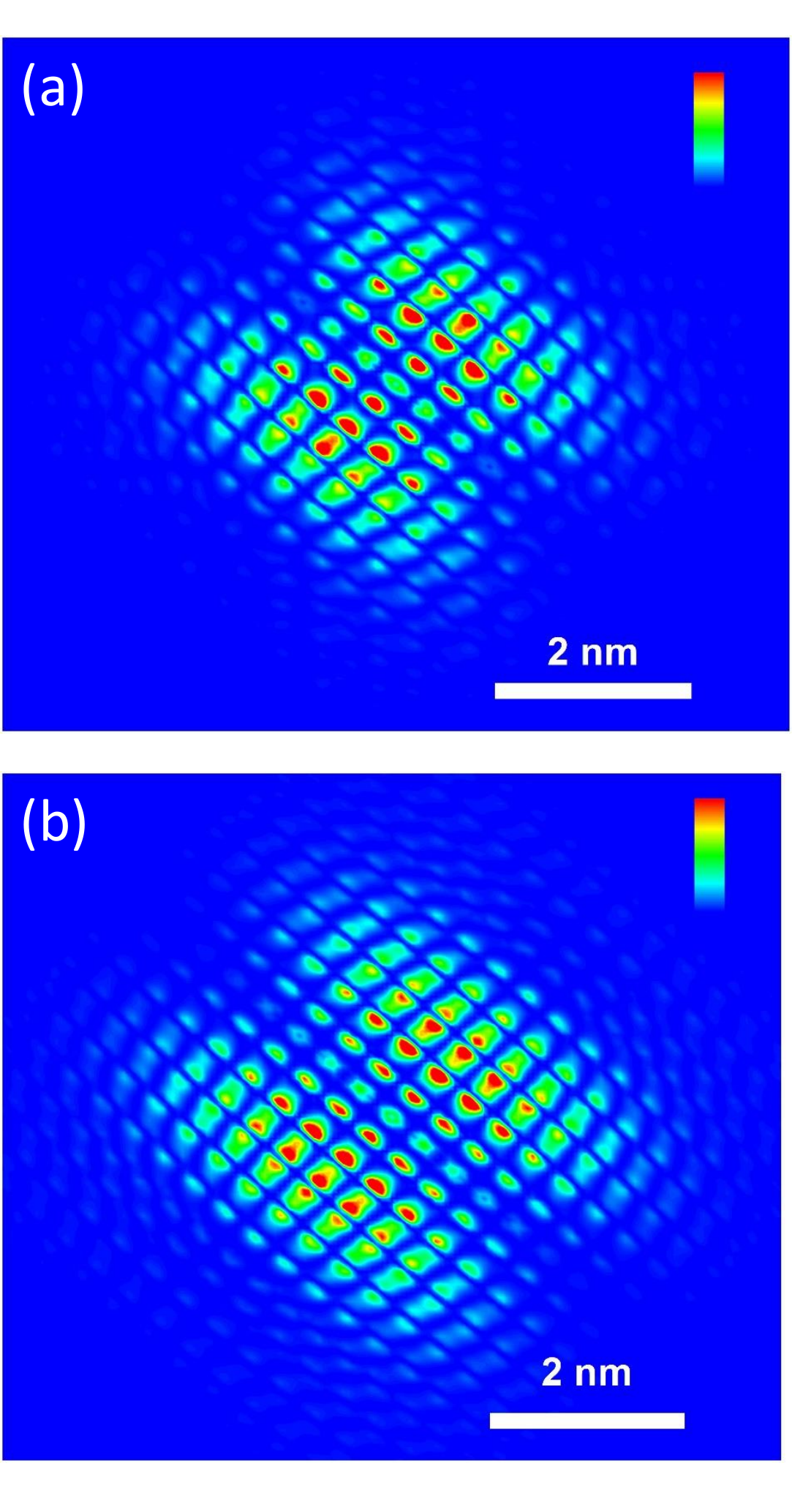}
\caption{Butterfly images of the charge distribution from donors at two equivalent (001) cuts but at different depths (a) $I^{(4)}_{3/4}$ and (b) $I^{(8)}_{3/4}$. The more distant donor (b) spreads over a larger area of the surface. The color scale ranges from 0 to (a) $4\times 10^{-4}(a_{\rm Si}^{-3})$ and (b) $10^{-6}\times (a_{\rm Si}^{-3})$.}
\label{fig:4vs8}
\end{figure}

We investigate to what extent  the image variety  presented here may provide information on the $(x,y,z)$ position of isolated substitutional donors in the Si lattice. If feasible, this would guide sorting favourable samples to perform specific spin qubit operations. For example, in a Si-donor-based quantum computer inspired by  Kane's original idea,~\cite{kaneNat1998} spin qubit operations are known to be highly sensitive to the donor position~\cite{koillerPRL2001}. In particular, two-qubit exchange gates are anticipated to require precise positioning of the interacting donor pairs.

All theoretical images have a center of inversion symmetry, which readily identifies the [100] and [010] (or $x$ and $y$) coordinates of the donor. This is not the case for the STM images, where patterns may or may not have a center of inversion symmetry (see Fig.~\ref{fig:EMT-vs-STM}(a) and (b) respectively). This aspect is related to the relative $(x,y)$ positioning of the donor with respect to the surface dimer rows, which may break one of the mirror symmetries (tranversal to the caterpillar figure) of the STM images. Properly taking into account this effect could lead to the donor $(x,y)$ position.

Laboratory fabrication of buried dopants in Si with atomically precise z-locations relative to the surface is a complicated issue at this early stage.  Uncertainties may be introduced by dopant diffusion during annealing.
{\it A posteriori} analysis of the shift of the conduction band edge can give a clue on the dopant depth with an error of the order of the lattice parameter.~\cite{salfiNatMat2014}

If the donor depth is determined by least-squares fitting of the band edge potential to an interval $[d-a_{\rm Si},d+a_{\rm Si}]$ containing 8 planes, with 95\% confidence, then the categorization into a B or C-type image reduces the number of possible planes to 4. 

Note that all features highlighted here are specific to $\langle 001 \rangle$ surfaces. In principle, the $\langle 111 \rangle$ surfaces discussed in Appendix~\ref{sec:111} form kaleidoscope-like figures that could be used to distinguish the donor depth uniquely.

Although all the results presented here correspond to a single donor, we may try to extract information about the relative position of two donors. A simple check for the same class of images is to compare the range of the charge distribution probed by STM. More distant donors correspond to more spread images, as illustrated in Fig.~\ref{fig:4vs8}. Identical images correspond theoretically to donors lying on the same $xy$ plane. However this is not a strict experimental requirement -- two donors with the same $z$ coordinates may appear differently at the surface due to the Si dimer rows  relative position with respect to each of them. Therefore, identification of donor pairs at the same depth below the surface requires careful consideration of the structural peculiarities of Si consistently combined with the reconstructed surface -- each donor's position relative to the surface dimers must be included in the analysis of the STM images. This feature may be included in our model, but the number of possible combinations do not fit into simple rules and should be analysed systematically in each case.

\section{Oscillatory behavior and donor-based qubits}
\label{sec:qubits}

It is possible to harness spins for quantum computation if we are able to control the spin-spin exchange coupling on demand. This task is difficult due to the sensitivity of the exchange coupling $J$ to the donors relative positioning~\cite{koillerPRB2004,koillerPRL2001,gonzalez2014}. Exchange coupling oscillations arise from the interference between the plane wave parts of {\em two donors} wavefunctions at a relative position  $\bf R$. Each donor establishes a pinning point for the 6  plane waves, leading to factors of the form $\exp [{\rm i}({\bf k}_\mu - {\bf k}_\nu)\cdot {\bf R}]$ in the exchange coupling. As a result  rapidly oscillating coupling $J({\bf R})$ is obtained along general directions ${\bf R}$.   There is however one favourable situation: donors kept at {\bf R} strictly along one of the $\langle 100 \rangle$ directions lead to a smoother behavior for $J({\bf R})$.

The fast oscillatory patterns of charge in our theoretical or experimental images are related not only to the plane wave part of the Bloch functions, but also show interferences coming from the periodic part of the Bloch functions. It is clear that the oscillatory behavior of the exchange coupling $J$ is not the same as the oscillatory pattern of the electronic wavefunction. The exchange oscillations are only due to the valley interference, since the periodic $u_\mu (\bf r)$ always interfere constructively.

\bibliography{bib-STM}

\begin{thebibliography}{26}%
\makeatletter
\providecommand \@ifxundefined [1]{%
 \@ifx{#1\undefined}
}%
\providecommand \@ifnum [1]{%
 \ifnum #1\expandafter \@firstoftwo
 \else \expandafter \@secondoftwo
 \fi
}%
\providecommand \@ifx [1]{%
 \ifx #1\expandafter \@firstoftwo
 \else \expandafter \@secondoftwo
 \fi
}%
\providecommand \natexlab [1]{#1}%
\providecommand \enquote  [1]{``#1''}%
\providecommand \bibnamefont  [1]{#1}%
\providecommand \bibfnamefont [1]{#1}%
\providecommand \citenamefont [1]{#1}%
\providecommand \href@noop [0]{\@secondoftwo}%
\providecommand \href [0]{\begingroup \@sanitize@url \@href}%
\providecommand \@href[1]{\@@startlink{#1}\@@href}%
\providecommand \@@href[1]{\endgroup#1\@@endlink}%
\providecommand \@sanitize@url [0]{\catcode `\\12\catcode `\$12\catcode
  `\&12\catcode `\#12\catcode `\^12\catcode `\_12\catcode `\%12\relax}%
\providecommand \@@startlink[1]{}%
\providecommand \@@endlink[0]{}%
\providecommand \url  [0]{\begingroup\@sanitize@url \@url }%
\providecommand \@url [1]{\endgroup\@href {#1}{\urlprefix }}%
\providecommand \urlprefix  [0]{URL }%
\providecommand \Eprint [0]{\href }%
\providecommand \doibase [0]{http://dx.doi.org/}%
\providecommand \selectlanguage [0]{\@gobble}%
\providecommand \bibinfo  [0]{\@secondoftwo}%
\providecommand \bibfield  [0]{\@secondoftwo}%
\providecommand \translation [1]{[#1]}%
\providecommand \BibitemOpen [0]{}%
\providecommand \bibitemStop [0]{}%
\providecommand \bibitemNoStop [0]{.\EOS\space}%
\providecommand \EOS [0]{\spacefactor3000\relax}%
\providecommand \BibitemShut  [1]{\csname bibitem#1\endcsname}%
\let\auto@bib@innerbib\@empty
\bibitem [{\citenamefont {Kohn}\ and\ \citenamefont
  {Luttinger}(1955)}]{kohnPR1955a}%
  \BibitemOpen
  \bibfield  {author} {\bibinfo {author} {\bibfnamefont {W.}~\bibnamefont
  {Kohn}}\ and\ \bibinfo {author} {\bibfnamefont {J.~M.}\ \bibnamefont
  {Luttinger}},\ }\href@noop {} {\bibfield  {journal} {\bibinfo  {journal}
  {Phys. Rev.}\ }\textbf {\bibinfo {volume} {98}},\ \bibinfo {pages} {915}
  (\bibinfo {year} {1955})}\BibitemShut {NoStop}%
\bibitem [{\citenamefont {Kittel}\ and\ \citenamefont
  {Mitchell}(1954)}]{kittel54}%
  \BibitemOpen
  \bibfield  {author} {\bibinfo {author} {\bibfnamefont {C.}~\bibnamefont
  {Kittel}}\ and\ \bibinfo {author} {\bibfnamefont {A.~H.}\ \bibnamefont
  {Mitchell}},\ }\href {\doibase 10.1103/PhysRev.96.1488} {\bibfield  {journal}
  {\bibinfo  {journal} {Phys. Rev.}\ }\textbf {\bibinfo {volume} {96}},\
  \bibinfo {pages} {1488} (\bibinfo {year} {1954})}\BibitemShut {NoStop}%
\bibitem [{\citenamefont {Zwanenburg}\ \emph {et~al.}(2013)\citenamefont
  {Zwanenburg}, \citenamefont {Dzurak}, \citenamefont {Morello}, \citenamefont
  {Simmons}, \citenamefont {Hollenberg}, \citenamefont {Klimeck}, \citenamefont
  {Rogge}, \citenamefont {Coppersmith},\ and\ \citenamefont
  {Eriksson}}]{zwanenburg2013}%
  \BibitemOpen
  \bibfield  {author} {\bibinfo {author} {\bibfnamefont {F.~A.}\ \bibnamefont
  {Zwanenburg}}, \bibinfo {author} {\bibfnamefont {A.~S.}\ \bibnamefont
  {Dzurak}}, \bibinfo {author} {\bibfnamefont {A.}~\bibnamefont {Morello}},
  \bibinfo {author} {\bibfnamefont {M.~Y.}\ \bibnamefont {Simmons}}, \bibinfo
  {author} {\bibfnamefont {L.~C.~L.}\ \bibnamefont {Hollenberg}}, \bibinfo
  {author} {\bibfnamefont {G.}~\bibnamefont {Klimeck}}, \bibinfo {author}
  {\bibfnamefont {S.}~\bibnamefont {Rogge}}, \bibinfo {author} {\bibfnamefont
  {S.~N.}\ \bibnamefont {Coppersmith}}, \ and\ \bibinfo {author} {\bibfnamefont
  {M.~A.}\ \bibnamefont {Eriksson}},\ }\href {\doibase
  10.1103/RevModPhys.85.961} {\bibfield  {journal} {\bibinfo  {journal} {Rev.
  Mod. Phys.}\ }\textbf {\bibinfo {volume} {85}},\ \bibinfo {pages} {961}
  (\bibinfo {year} {2013})}\BibitemShut {NoStop}%
\bibitem [{\citenamefont {Salfi}\ \emph {et~al.}(2014)\citenamefont {Salfi},
  \citenamefont {Mol}, \citenamefont {Rahman}, \citenamefont {Klimeck},
  \citenamefont {Simmons}, \citenamefont {Hollenberg},\ and\ \citenamefont
  {Rogge}}]{salfiNatMat2014}%
  \BibitemOpen
  \bibfield  {author} {\bibinfo {author} {\bibfnamefont {J.}~\bibnamefont
  {Salfi}}, \bibinfo {author} {\bibfnamefont {J.~A.}\ \bibnamefont {Mol}},
  \bibinfo {author} {\bibfnamefont {R.}~\bibnamefont {Rahman}}, \bibinfo
  {author} {\bibfnamefont {G.}~\bibnamefont {Klimeck}}, \bibinfo {author}
  {\bibfnamefont {M.~Y.}\ \bibnamefont {Simmons}}, \bibinfo {author}
  {\bibfnamefont {L.~C.~L.}\ \bibnamefont {Hollenberg}}, \ and\ \bibinfo
  {author} {\bibfnamefont {S.}~\bibnamefont {Rogge}},\ }\href@noop {}
  {\bibfield  {journal} {\bibinfo  {journal} {Nat Mater}\ }\textbf {\bibinfo
  {volume} {13}},\ \bibinfo {pages} {605} (\bibinfo {year} {2014})}\BibitemShut
  {NoStop}%
\bibitem [{\citenamefont {Hensel}\ \emph {et~al.}(1965)\citenamefont {Hensel},
  \citenamefont {Hasegawa},\ and\ \citenamefont {Nakayama}}]{hensel1965}%
  \BibitemOpen
  \bibfield  {author} {\bibinfo {author} {\bibfnamefont {J.~C.}\ \bibnamefont
  {Hensel}}, \bibinfo {author} {\bibfnamefont {H.}~\bibnamefont {Hasegawa}}, \
  and\ \bibinfo {author} {\bibfnamefont {M.}~\bibnamefont {Nakayama}},\ }\href
  {\doibase 10.1103/PhysRev.138.A225} {\bibfield  {journal} {\bibinfo
  {journal} {Phys. Rev.}\ }\textbf {\bibinfo {volume} {138}},\ \bibinfo {pages}
  {A225} (\bibinfo {year} {1965})}\BibitemShut {NoStop}%
\bibitem [{\citenamefont {Feher}(1959)}]{feher1959}%
  \BibitemOpen
  \bibfield  {author} {\bibinfo {author} {\bibfnamefont {G.}~\bibnamefont
  {Feher}},\ }\href {\doibase 10.1103/PhysRev.114.1219} {\bibfield  {journal}
  {\bibinfo  {journal} {Phys. Rev.}\ }\textbf {\bibinfo {volume} {114}},\
  \bibinfo {pages} {1219} (\bibinfo {year} {1959})}\BibitemShut {NoStop}%
\bibitem [{\citenamefont {Saraiva}\ \emph {et~al.}(2011)\citenamefont
  {Saraiva}, \citenamefont {Calder\'on}, \citenamefont {Capaz}, \citenamefont
  {Hu}, \citenamefont {Das~Sarma},\ and\ \citenamefont
  {Koiller}}]{saraiva2011}%
  \BibitemOpen
  \bibfield  {author} {\bibinfo {author} {\bibfnamefont {A.~L.}\ \bibnamefont
  {Saraiva}}, \bibinfo {author} {\bibfnamefont {M.~J.}\ \bibnamefont
  {Calder\'on}}, \bibinfo {author} {\bibfnamefont {R.~B.}\ \bibnamefont
  {Capaz}}, \bibinfo {author} {\bibfnamefont {X.}~\bibnamefont {Hu}}, \bibinfo
  {author} {\bibfnamefont {S.}~\bibnamefont {Das~Sarma}}, \ and\ \bibinfo
  {author} {\bibfnamefont {B.}~\bibnamefont {Koiller}},\ }\href {\doibase
  10.1103/PhysRevB.84.155320} {\bibfield  {journal} {\bibinfo  {journal} {Phys.
  Rev. B}\ }\textbf {\bibinfo {volume} {84}},\ \bibinfo {pages} {155320}
  (\bibinfo {year} {2011})}\BibitemShut {NoStop}%
\bibitem [{\citenamefont {Saraiva}\ \emph {et~al.}(2015)\citenamefont
  {Saraiva}, \citenamefont {Baena}, \citenamefont {Calder\'on},\ and\
  \citenamefont {Koiller}}]{saraivaJPCM2015}%
  \BibitemOpen
  \bibfield  {author} {\bibinfo {author} {\bibfnamefont {A.~L.}\ \bibnamefont
  {Saraiva}}, \bibinfo {author} {\bibfnamefont {A.}~\bibnamefont {Baena}},
  \bibinfo {author} {\bibfnamefont {M.~J.}\ \bibnamefont {Calder\'on}}, \ and\
  \bibinfo {author} {\bibfnamefont {B.}~\bibnamefont {Koiller}},\ }\href@noop
  {} {\bibfield  {journal} {\bibinfo  {journal} {Journal of Physics: Condensed
  Matter}\ }\textbf {\bibinfo {volume} {27}},\ \bibinfo {pages} {154208}
  (\bibinfo {year} {2015})}\BibitemShut {NoStop}%
\bibitem [{\citenamefont {Tersoff}\ and\ \citenamefont
  {Hamann}(1985)}]{tersoffPRB1985}%
  \BibitemOpen
  \bibfield  {author} {\bibinfo {author} {\bibfnamefont {J.}~\bibnamefont
  {Tersoff}}\ and\ \bibinfo {author} {\bibfnamefont {D.~R.}\ \bibnamefont
  {Hamann}},\ }\href {\doibase 10.1103/PhysRevB.31.805} {\bibfield  {journal}
  {\bibinfo  {journal} {Phys. Rev. B}\ }\textbf {\bibinfo {volume} {31}},\
  \bibinfo {pages} {805} (\bibinfo {year} {1985})}\BibitemShut {NoStop}%
\bibitem [{\citenamefont {Miwa}\ \emph {et~al.}(2013)\citenamefont {Miwa},
  \citenamefont {Mol}, \citenamefont {Salfi}, \citenamefont {Rogge},\ and\
  \citenamefont {Simmons}}]{miwaAPL2013}%
  \BibitemOpen
  \bibfield  {author} {\bibinfo {author} {\bibfnamefont {J.~A.}\ \bibnamefont
  {Miwa}}, \bibinfo {author} {\bibfnamefont {J.~A.}\ \bibnamefont {Mol}},
  \bibinfo {author} {\bibfnamefont {J.}~\bibnamefont {Salfi}}, \bibinfo
  {author} {\bibfnamefont {S.~S.}\ \bibnamefont {Rogge}}, \ and\ \bibinfo
  {author} {\bibfnamefont {M.~Y.}\ \bibnamefont {Simmons}},\ }\href@noop {}
  {\bibfield  {journal} {\bibinfo  {journal} {Applied Physics Letters}\
  }\textbf {\bibinfo {volume} {103}},\ \bibinfo {pages} {043106} (\bibinfo
  {year} {2013})}\BibitemShut {NoStop}%
\bibitem [{\citenamefont {Voisin}\ \emph {et~al.}(2015)\citenamefont {Voisin},
  \citenamefont {Salfi}, \citenamefont {Bocquel}, \citenamefont {Rahman},\ and\
  \citenamefont {Rogge}}]{voisinJPCM2015l}%
  \BibitemOpen
  \bibfield  {author} {\bibinfo {author} {\bibfnamefont {B.}~\bibnamefont
  {Voisin}}, \bibinfo {author} {\bibfnamefont {J.}~\bibnamefont {Salfi}},
  \bibinfo {author} {\bibfnamefont {J.}~\bibnamefont {Bocquel}}, \bibinfo
  {author} {\bibfnamefont {R.}~\bibnamefont {Rahman}}, \ and\ \bibinfo {author}
  {\bibfnamefont {S.~S.}\ \bibnamefont {Rogge}},\ }\href@noop {} {\bibfield
  {journal} {\bibinfo  {journal} {J. Phys.: Condens. Matter}\ }\textbf
  {\bibinfo {volume} {27}},\ \bibinfo {pages} {154203} (\bibinfo {year}
  {2015})}\BibitemShut {NoStop}%
\bibitem [{\citenamefont {{Mol, J A}}\ \emph {et~al.}(2013)\citenamefont {{Mol,
  J A}}, \citenamefont {Salfi}, \citenamefont {Miwa}, \citenamefont {Simmons},\
  and\ \citenamefont {Rogge}}]{molPRB2013}%
  \BibitemOpen
  \bibfield  {author} {\bibinfo {author} {\bibnamefont {{Mol, J A}}}, \bibinfo
  {author} {\bibfnamefont {J.}~\bibnamefont {Salfi}}, \bibinfo {author}
  {\bibfnamefont {J.~A.}\ \bibnamefont {Miwa}}, \bibinfo {author}
  {\bibfnamefont {M.~Y.}\ \bibnamefont {Simmons}}, \ and\ \bibinfo {author}
  {\bibfnamefont {S.}~\bibnamefont {Rogge}},\ }\href@noop {} {\bibfield
  {journal} {\bibinfo  {journal} {Phys. Rev. B}\ }\textbf {\bibinfo {volume}
  {87}},\ \bibinfo {pages} {245417} (\bibinfo {year} {2013})}\BibitemShut
  {NoStop}%
\bibitem [{\citenamefont {Koiller}\ \emph {et~al.}(2004)\citenamefont
  {Koiller}, \citenamefont {Capaz}, \citenamefont {Hu},\ and\ \citenamefont
  {Das~Sarma}}]{koillerPRB2004}%
  \BibitemOpen
  \bibfield  {author} {\bibinfo {author} {\bibfnamefont {B.}~\bibnamefont
  {Koiller}}, \bibinfo {author} {\bibfnamefont {R.~B.}\ \bibnamefont {Capaz}},
  \bibinfo {author} {\bibfnamefont {X.}~\bibnamefont {Hu}}, \ and\ \bibinfo
  {author} {\bibfnamefont {S.}~\bibnamefont {Das~Sarma}},\ }\href {\doibase
  10.1103/PhysRevB.70.115207} {\bibfield  {journal} {\bibinfo  {journal} {Phys.
  Rev. B}\ }\textbf {\bibinfo {volume} {70}},\ \bibinfo {pages} {115207}
  (\bibinfo {year} {2004})}\BibitemShut {NoStop}%
\bibitem [{\citenamefont {Yakunin}\ \emph {et~al.}(2004)\citenamefont
  {Yakunin}, \citenamefont {Silov}, \citenamefont {Koenraad}, \citenamefont
  {Wolter}, \citenamefont {Van~Roy}, \citenamefont {De~Boeck}, \citenamefont
  {Tang},\ and\ \citenamefont {Flatt\'e}}]{yakuninPRL2004}%
  \BibitemOpen
  \bibfield  {author} {\bibinfo {author} {\bibfnamefont {A.~M.}\ \bibnamefont
  {Yakunin}}, \bibinfo {author} {\bibfnamefont {A.~Y.}\ \bibnamefont {Silov}},
  \bibinfo {author} {\bibfnamefont {P.~M.}\ \bibnamefont {Koenraad}}, \bibinfo
  {author} {\bibfnamefont {J.~H.}\ \bibnamefont {Wolter}}, \bibinfo {author}
  {\bibfnamefont {W.}~\bibnamefont {Van~Roy}}, \bibinfo {author} {\bibfnamefont
  {J.}~\bibnamefont {De~Boeck}}, \bibinfo {author} {\bibfnamefont {J.-M.}\
  \bibnamefont {Tang}}, \ and\ \bibinfo {author} {\bibfnamefont {M.~E.}\
  \bibnamefont {Flatt\'e}},\ }\href {\doibase 10.1103/PhysRevLett.92.216806}
  {\bibfield  {journal} {\bibinfo  {journal} {Phys. Rev. Lett.}\ }\textbf
  {\bibinfo {volume} {92}},\ \bibinfo {pages} {216806} (\bibinfo {year}
  {2004})}\BibitemShut {NoStop}%
\bibitem [{\citenamefont {Mol}\ \emph {et~al.}(2015)\citenamefont {Mol},
  \citenamefont {Salfi}, \citenamefont {Rahman}, \citenamefont {Hsueh},
  \citenamefont {Miwa}, \citenamefont {Klimeck}, \citenamefont {Simmons},\ and\
  \citenamefont {Rogge}}]{molAPL2015}%
  \BibitemOpen
  \bibfield  {author} {\bibinfo {author} {\bibfnamefont {J.~A.}\ \bibnamefont
  {Mol}}, \bibinfo {author} {\bibfnamefont {J.}~\bibnamefont {Salfi}}, \bibinfo
  {author} {\bibfnamefont {R.}~\bibnamefont {Rahman}}, \bibinfo {author}
  {\bibfnamefont {Y.}~\bibnamefont {Hsueh}}, \bibinfo {author} {\bibfnamefont
  {J.~A.}\ \bibnamefont {Miwa}}, \bibinfo {author} {\bibfnamefont
  {G.}~\bibnamefont {Klimeck}}, \bibinfo {author} {\bibfnamefont {M.~Y.}\
  \bibnamefont {Simmons}}, \ and\ \bibinfo {author} {\bibfnamefont {S.~S.}\
  \bibnamefont {Rogge}},\ }\href@noop {} {\bibfield  {journal} {\bibinfo
  {journal} {Applied Physics Letters}\ }\textbf {\bibinfo {volume} {106}},\
  \bibinfo {pages} {203110} (\bibinfo {year} {2015})}\BibitemShut {NoStop}%
\bibitem [{\citenamefont {Koiller}\ \emph {et~al.}(2001)\citenamefont
  {Koiller}, \citenamefont {Hu},\ and\ \citenamefont
  {Das~Sarma}}]{koillerPRL2001}%
  \BibitemOpen
  \bibfield  {author} {\bibinfo {author} {\bibfnamefont {B.}~\bibnamefont
  {Koiller}}, \bibinfo {author} {\bibfnamefont {X.}~\bibnamefont {Hu}}, \ and\
  \bibinfo {author} {\bibfnamefont {S.}~\bibnamefont {Das~Sarma}},\ }\href
  {\doibase 10.1103/PhysRevLett.88.027903} {\bibfield  {journal} {\bibinfo
  {journal} {Phys. Rev. Lett.}\ }\textbf {\bibinfo {volume} {88}},\ \bibinfo
  {pages} {027903} (\bibinfo {year} {2001})}\BibitemShut {NoStop}%
\bibitem [{\citenamefont {Gamble}\ \emph {et~al.}(2015)\citenamefont {Gamble},
  \citenamefont {Jacobson}, \citenamefont {Nielsen}, \citenamefont {Baczewski},
  \citenamefont {Moussa}, \citenamefont {Monta\~no},\ and\ \citenamefont
  {Muller}}]{gamble2015}%
  \BibitemOpen
  \bibfield  {author} {\bibinfo {author} {\bibfnamefont {J.~K.}\ \bibnamefont
  {Gamble}}, \bibinfo {author} {\bibfnamefont {N.~T.}\ \bibnamefont
  {Jacobson}}, \bibinfo {author} {\bibfnamefont {E.}~\bibnamefont {Nielsen}},
  \bibinfo {author} {\bibfnamefont {A.~D.}\ \bibnamefont {Baczewski}}, \bibinfo
  {author} {\bibfnamefont {J.~E.}\ \bibnamefont {Moussa}}, \bibinfo {author}
  {\bibfnamefont {I.}~\bibnamefont {Monta\~no}}, \ and\ \bibinfo {author}
  {\bibfnamefont {R.~P.}\ \bibnamefont {Muller}},\ }\href {\doibase
  10.1103/PhysRevB.91.235318} {\bibfield  {journal} {\bibinfo  {journal} {Phys.
  Rev. B}\ }\textbf {\bibinfo {volume} {91}},\ \bibinfo {pages} {235318}
  (\bibinfo {year} {2015})}\BibitemShut {NoStop}%
\bibitem [{\citenamefont {Gonzalez-Zalba}\ \emph {et~al.}(2014)\citenamefont
  {Gonzalez-Zalba}, \citenamefont {Saraiva}, \citenamefont {Calderón},
  \citenamefont {Heiss}, \citenamefont {Koiller},\ and\ \citenamefont
  {Ferguson}}]{gonzalez2014}%
  \BibitemOpen
  \bibfield  {author} {\bibinfo {author} {\bibfnamefont {M.~F.}\ \bibnamefont
  {Gonzalez-Zalba}}, \bibinfo {author} {\bibfnamefont {A.}~\bibnamefont
  {Saraiva}}, \bibinfo {author} {\bibfnamefont {M.~J.}\ \bibnamefont
  {Calderón}}, \bibinfo {author} {\bibfnamefont {D.}~\bibnamefont {Heiss}},
  \bibinfo {author} {\bibfnamefont {B.}~\bibnamefont {Koiller}}, \ and\
  \bibinfo {author} {\bibfnamefont {A.~J.}\ \bibnamefont {Ferguson}},\ }\href
  {\doibase 10.1021/nl5023942} {\bibfield  {journal} {\bibinfo  {journal} {Nano
  Letters}\ }\textbf {\bibinfo {volume} {14}},\ \bibinfo {pages} {5672}
  (\bibinfo {year} {2014})},\ \bibinfo {note} {pMID: 25230333}\BibitemShut
  {NoStop}%
\bibitem [{\citenamefont {Macfarlane}\ and\ \citenamefont
  {Roberts}(1955)}]{macfarlanePR1955}%
  \BibitemOpen
  \bibfield  {author} {\bibinfo {author} {\bibfnamefont {G.~G.}\ \bibnamefont
  {Macfarlane}}\ and\ \bibinfo {author} {\bibfnamefont {V.}~\bibnamefont
  {Roberts}},\ }\href {\doibase 10.1103/PhysRev.98.1865} {\bibfield  {journal}
  {\bibinfo  {journal} {Phys. Rev.}\ }\textbf {\bibinfo {volume} {98}},\
  \bibinfo {pages} {1865} (\bibinfo {year} {1955})}\BibitemShut {NoStop}%
\bibitem [{\citenamefont {Baldereschi}(1970)}]{baldereschiPRB1970}%
  \BibitemOpen
  \bibfield  {author} {\bibinfo {author} {\bibfnamefont {A.}~\bibnamefont
  {Baldereschi}},\ }\href {\doibase 10.1103/PhysRevB.1.4673} {\bibfield
  {journal} {\bibinfo  {journal} {Phys. Rev. B}\ }\textbf {\bibinfo {volume}
  {1}},\ \bibinfo {pages} {4673} (\bibinfo {year} {1970})}\BibitemShut
  {NoStop}%
\bibitem [{\citenamefont {Boland}(1990)}]{bolandPRL1990}%
  \BibitemOpen
  \bibfield  {author} {\bibinfo {author} {\bibfnamefont {J.}~\bibnamefont
  {Boland}},\ }\href@noop {} {\bibfield  {journal} {\bibinfo  {journal} {Phys.
  Rev. Lett.}\ }\textbf {\bibinfo {volume} {65}},\ \bibinfo {pages} {3325}
  (\bibinfo {year} {1990})}\BibitemShut {NoStop}%
\bibitem [{\citenamefont {Oura}\ \emph {et~al.}(1999)\citenamefont {Oura},
  \citenamefont {Lifshits}, \citenamefont {Saranin}, \citenamefont {Zotov},\
  and\ \citenamefont {Katayama}}]{ouraSurfSciRep1999}%
  \BibitemOpen
  \bibfield  {author} {\bibinfo {author} {\bibfnamefont {K.}~\bibnamefont
  {Oura}}, \bibinfo {author} {\bibfnamefont {V.~G.}\ \bibnamefont {Lifshits}},
  \bibinfo {author} {\bibfnamefont {A.~A.}\ \bibnamefont {Saranin}}, \bibinfo
  {author} {\bibfnamefont {A.~V.}\ \bibnamefont {Zotov}}, \ and\ \bibinfo
  {author} {\bibfnamefont {M.}~\bibnamefont {Katayama}},\ }\href@noop {}
  {\bibfield  {journal} {\bibinfo  {journal} {Surface Science Reports}\
  }\textbf {\bibinfo {volume} {35}},\ \bibinfo {pages} {1} (\bibinfo {year}
  {1999})}\BibitemShut {NoStop}%
\bibitem [{\citenamefont {Dagata}\ \emph {et~al.}(1990)\citenamefont {Dagata},
  \citenamefont {Schneir}, \citenamefont {Harary}, \citenamefont {Evans},
  \citenamefont {Postek},\ and\ \citenamefont {Bennett}}]{dagataAPL1990}%
  \BibitemOpen
  \bibfield  {author} {\bibinfo {author} {\bibfnamefont {J.~A.}\ \bibnamefont
  {Dagata}}, \bibinfo {author} {\bibfnamefont {J.}~\bibnamefont {Schneir}},
  \bibinfo {author} {\bibfnamefont {H.~H.}\ \bibnamefont {Harary}}, \bibinfo
  {author} {\bibfnamefont {C.~J.}\ \bibnamefont {Evans}}, \bibinfo {author}
  {\bibfnamefont {M.~T.}\ \bibnamefont {Postek}}, \ and\ \bibinfo {author}
  {\bibfnamefont {J.}~\bibnamefont {Bennett}},\ }\href@noop {} {\bibfield
  {journal} {\bibinfo  {journal} {Applied Physics Letters}\ }\textbf {\bibinfo
  {volume} {56}},\ \bibinfo {pages} {2001} (\bibinfo {year}
  {1990})}\BibitemShut {NoStop}%
\bibitem [{\citenamefont {Higashi}\ \emph {et~al.}(1990)\citenamefont
  {Higashi}, \citenamefont {Chabal}, \citenamefont {Trucks},\ and\
  \citenamefont {Raghavachari}}]{higashiAPL1990}%
  \BibitemOpen
  \bibfield  {author} {\bibinfo {author} {\bibfnamefont {G.~S.}\ \bibnamefont
  {Higashi}}, \bibinfo {author} {\bibfnamefont {Y.~J.}\ \bibnamefont {Chabal}},
  \bibinfo {author} {\bibfnamefont {G.~W.}\ \bibnamefont {Trucks}}, \ and\
  \bibinfo {author} {\bibfnamefont {K.}~\bibnamefont {Raghavachari}},\
  }\href@noop {} {\bibfield  {journal} {\bibinfo  {journal} {Appl. Phys.
  Lett.}\ }\textbf {\bibinfo {volume} {56}},\ \bibinfo {pages} {656} (\bibinfo
  {year} {1990})}\BibitemShut {NoStop}%
\bibitem [{\citenamefont {Owman}\ and\ \citenamefont
  {M{\aa}rtensson}(1994)}]{owmanSurfSci1994}%
  \BibitemOpen
  \bibfield  {author} {\bibinfo {author} {\bibfnamefont {F.}~\bibnamefont
  {Owman}}\ and\ \bibinfo {author} {\bibfnamefont {P.}~\bibnamefont
  {M{\aa}rtensson}},\ }\href@noop {} {\bibfield  {journal} {\bibinfo  {journal}
  {Surface Science}\ }\textbf {\bibinfo {volume} {303}},\ \bibinfo {pages}
  {L367} (\bibinfo {year} {1994})}\BibitemShut {NoStop}%
\bibitem [{\citenamefont {Kane}(1998)}]{kaneNat1998}%
  \BibitemOpen
  \bibfield  {author} {\bibinfo {author} {\bibfnamefont {B.~E.}\ \bibnamefont
  {Kane}},\ }\href@noop {} {\bibfield  {journal} {\bibinfo  {journal} {Nature}\
  }\textbf {\bibinfo {volume} {393}},\ \bibinfo {pages} {133} (\bibinfo {year}
  {1998})}\BibitemShut {NoStop}%
\end{thebibliography}%
\end{document}